\documentclass[]{interact}

\usepackage{epstopdf}
\usepackage[caption=false]{subfig}

\usepackage[numbers,sort&compress]{natbib}
\bibpunct[, ]{[}{]}{,}{n}{,}{,}
\renewcommand\bibfont{\fontsize{10}{12}\selectfont}
\makeatletter
\def\NAT@def@citea{\def\@citea{\NAT@separator}}
\makeatother

\theoremstyle{plain}
\newtheorem{theorem}{Theorem}[section]
\newtheorem{lemma}[theorem]{Lemma}

\newtheorem{proposition}[theorem]{Proposition}

\theoremstyle{definition}

\theoremstyle{remark}
\newtheorem{remark}{Remark}

\def\RR{{\mathbb R}}
\usepackage{algpseudocode}
\usepackage{algorithm}
\usepackage{hyperref}
\usepackage{amsmath}
\usepackage{amssymb}
\usepackage{mathtools}
\usepackage{multirow}
\usepackage{siunitx}

\begin{document}

\articletype{ARTICLE TEMPLATE}

\title{Exploring dimension learning via a penalized probabilistic principal component analysis}

\author{
\name{Wei Q. Deng\textsuperscript{a, b}\thanks{CONTACT W. Q. Deng. Email: dengwq@mcmaster.ca} and Radu V. Craiu\textsuperscript{c}}
\affil{\textsuperscript{a} Department of Psychiatry and Behavioural Neurosciences, McMaster University\\
\textsuperscript{b} Peter Boris Centre for Addictions Research, St. Joseph's Healthcare Hamilton, L8P 3R2, Canada\\ 
\textsuperscript{c} Department of Statistical Sciences, University of Toronto, Toronto, M5S 3G3, Canada}
}

\maketitle

\begin{abstract}
Establishing a low-dimensional representation of the data leads to efficient data learning strategies. In many cases, the reduced dimension needs to be explicitly stated and estimated from the data. We explore the estimation of dimension in finite samples as a constrained optimization problem, where the estimated dimension is a maximizer of a penalized profile likelihood criterion within the framework of a probabilistic principal components analysis. Unlike other penalized maximization problems that require an ``optimal'' penalty tuning parameter, we propose a data-averaging procedure whereby the estimated dimension emerges as the most favourable choice over a range of plausible penalty parameters. The proposed heuristic is compared to a large number of alternative criteria in simulations and an application to gene expression data. Extensive simulation studies reveal that none of the methods uniformly dominate the other and highlight the importance of subject-specific knowledge in choosing statistical methods for dimension learning. Our application results also suggest that gene expression data have a higher intrinsic dimension than previously thought. Overall, our proposed heuristic strikes a good balance and is the method of choice when model assumptions deviated moderately.
\end{abstract}

\begin{keywords}
dimension estimation; model selection; penalization; principal component analysis; probabilistic principal component analysis; profile likelihood.
\end{keywords}

\section{Introduction}

Consider a data matrix $X \in \mathbb{R}^{n \times m}$ that has been column and row centered such that
\[
\sum_{i = 1}^n x_{ij} = \sum_{j = 1}^m x_{ij} = 0; \quad \text{for } i = 1, \dots, n; \text{ and } j = 1, \dots, m, 
\]
we are interested in a linear decomposition of $X$ to a signal component, driven by variance in the top singular values, and a noise component of the form:
\begin{equation}
X = WL + F, \label{eqn:X}
\end{equation}
where $W \in \mathbb{R}^{n \times k}$ is a constant matrix with rank $k < n$, $L\in \mathbb{R}^{k \times m}$ is an arbitrary matrix with orthonormal columns, and $F$ is a matrix whose rows are uncorrelated and have equal variance. The dimension of interest depends on $W$ as it is the minimal rank $k$ such that rows of $X - WL$ are uncorrelated and have isotropic covariance. Henceforth, we refer to $k$ as the \textit{effective rank} of the data because, intuitively, correlation structure in the rows of $X$ reduces data dimension attributed to the signal component ($WL$) from $\text{min}(n, m)$ to $k$.  

Estimation of $k$ has been studied in various contexts as the linear model~\eqref{eqn:X} has many alternative forms and names, such as a principal component analysis (PCA;~\citealp{pearson1901liii,hotelling1992relations,hotelling1933analysis}),
a truncated singular value decomposition (SVD), a factor analysis model~\citep{bartholomew1987latent}, and a spiked population model~\citep{johnstone2001distribution}, where the effective rank coincides with the definition of the number of spikes. 

The approaches to determine $k$ as the number of principal components (PCs), can be summarized under roughly three categories according to Jolliffe~\citep{jolliffe2002choosing}. The first type is a variety of \textit{ad-hoc} rules that have an empirical basis, such as the scree test~\citep{cattell1966scree} or Kaiser rule. To automate the decision, Zhu and Ghodsi proposed a profile likelihood criterion that detects a ``gap'' in the sample eigenvalues~\cite{zhu2006automatic}. A second class of methods rely on asymptotic tests, such as the likelihood ratio test for equality of eigenvalues \citep{bartlett1954note, lawley1956tests,ledoit2002,SCHOTT2006827,FORZANI201718}, which differ according to asymptotic conditions on the data dimensions. Instead of an asymptotic test, Choi et al. \cite{choi2017selecting} recently proposed an exact method for hypothesis testing of signals in a noisy matrix to estimate the number of PCs that showed promising results in simulations. Finally, for small datasets, computational methods such as bootstrap, permutation and cross-validation can be implemented in a timely manner. Among them, cross-validation is frequently used~\citep{mardia1979multivariate} with a general cross-validation (GCV) criterion~\cite{josse2012selecting} that also works well with large datasets.

Using a truncated SVD, Gavish and Donoho~\cite{gavish2014optimal} proposed to remove the underlying noise in the singular values via a hard threshold-based approach. In this case, the stopping rule based on a single threshold could be useful for recovering the original data in the sense of asymptotic mean squared error, but does not directly inform the minimal rank of the noise reduced data. Similarly in isotropic factor analysis, Bai and Ng~\citep{Bai2002} proposed to estimate the number of factors by finding some threshold to separate large and small eigenvalues of the data covariance matrix that leverages various penalty functions, but the approach depends on the correct estimation of error variance. Using a different strategy, Passemier \textit{et al}.~\cite{passemier2017estimation} tackled the estimation of the noise variance, which led to a bias-corrected criterion for estimating $k$ when $n >> m$. 

Here we focus on reviewing model-based methods where the solution arises from various model selection criteria. Probabilistic principal component analysis (PPCA), introduced in the seminal paper of Tipping and Bishop~\cite{Tipping1999a}, allows the estimation of $k$ as a likelihood optimization problem. An alternative Bayesian approach was proposed in~\citep{Bishop19991}, with the caveat that the full Bayesian estimation using Markov Chain Monte Carlo can be computationally prohibitive for large datasets~\citep{hoff2007model} and approximations are needed. Indeed, Minka implemented Laplace's method to approximate the posterior likelihood~\cite{minka2001automatic} and showed it to be often superior to cross-validation and variational inference \citep{Bishop19992, nakajima2012perfect} with the added benefit of fast computation. An exact marginal likelihood criterion based on a normal-gamma prior distribution has been developed that is competitive with both Bayesian and frequentist methods in low dimensional settings \citep{bouveyron2017exact}. For high-dimensional data with a small number of observations, Hoyle \cite{hoyle2008automatic} noted the unsatisfactory performance of Laplace's approximation and proposed to modify the Bayesian model using a Gaussian parametrization that showed improved performance. Observing the symmetry in the data structure, Sobczyk et al.~\cite{sobczyk2017bayesian} approximated the Bayesian models for both $X$ and $X^{T}$, and thus proposed two separate criteria that work well under divergence of either the number of observations ($m$) or samples ($n$), while the other one is constant.

Penalized maximum likelihood approaches are widely used to induce sparsity in the number of parameters used to characterize statistical models and have proven suitable for model selection. Here we explore using penalized probabilistic PCA models to estimate the effective rank and propose an accompanying data-driven heuristic to estimate the dimension. This heuristic has theoretical basis, was examined in extensive simulations and applied to a microarray gene expression dataset to inform the data dimension. We find that the penalized approach is competitive when compared to Bayesian and empirical alternatives in both simulated and application data, especially under departure from independence and normality assumptions. None of the methods uniformly dominate the others across the wide range of conditions, highlighting the importance of verifying the assumptions underlying each method.

This paper is structured as follows. We first revisit the probabilistic principal component model in Section~\ref{s:ppca}. In Section~\ref{s:pppca}, we explore using the penalized probabilistic PCA to model the data dimension as part of the optimization problem and present a data-driven algorithm for dimension learning. Results from an extensive simulation study comparing different classes of methods are presented in Section~\ref{s:sim_res} and an application to gene expression data is presented in Section~\ref{s:app_geneExp}. In the last section, we conclude the paper with general remarks on the proposed penalized approach and our practical recommendation to dimension learning in data applications.

\section{Probabilistic principal components analysis}
\label{s:ppca}

Given data $X \in \mathbb{R}^{n\times m}$, we seek a low-dimensional representation in the columns of $X$,  $X_j \in \mathbb{R}^{n}$, $j = 1, \dots, m$. Suppose there exists a fixed dimension $k \in \mathrm{Z}_+ (1 \le k \le n-1)$ such that:
\begin{equation}
X_j = \mu + {W}{l_j} + f_j, \quad j=1,\dots, m\label{eq:generative_model}
\end{equation}
where $\mu$ is the mean vector, ${W} \in \mathbb{R}^{n \times k}$ is a constant matrix, $l_j \in \mathbb{R}^{k}$ is a latent vector, and $f_j \in \mathbb{R}^{n}$ is noise in the data. In order to identify the data decomposition to signal ($WL$) and noise ($F$) components, we make the assumption that both the latent vector and the noise component are spherical Gaussian. This decomposition implies that the $n$-dimensional vector $X_j$ is obtained as a linear transformation of a $k$-dimensional latent vector. Therefore, the spanned subspace of $X_1,\ldots, X_m$ has effective dimension $k$. The value of $k$ is unknown in realistic examples and needs to be estimated from the data $X$. The usual PCA decomposition is obtained when the dimension is $k=n$, and in this case, $F$ in equation~\eqref{eqn:X} reduces to $\mathbf{0}$.

In this paper, we assume $f_j \sim \mathcal{N}(0, \zeta^2I_{n})$ and $l_j \sim \mathcal{N}(0,I_k)$, which imply that for any $1\le j \le m$, $X_j$ follows the Gaussian  distribution:
\begin{equation}
X_j \sim \mathcal{N}(\mu, WW^{T} + \zeta^2I_{n}).
\label{eq:distX}
\end{equation}

Denote the covariance matrix of $X_j$ by $\Phi = WW^{T} + \zeta^2I_{n}$ and under model~\eqref{eq:distX} it has a maximum of $k+1$ unique eigenvalues: $\lambda_1, \dots, \lambda_k$ and $\zeta^2$. This model forces the samples, represented by the rows of X, to be conditionally independent given the random vectors, $\{l_1,\ldots,l_k\}$, and thus the covariance matrix $WW^{T} + \zeta^2I_{n}$ can take on a more parsimonious representation. In general, the latent vectors may not have a Gaussian distribution and can be used to specify non-Gaussian signal components, such as those in a linear noisy independent component analysis model.

The log-likelihood function with respect to the unknown parameters $W$ and $\zeta^2$, given independent observations $X = (x_1, \dots, x_m)$, is denoted by
\begin{equation}
l(\mu, W,\zeta^{2}; X)=\frac{-m}{2}[n\log(2\pi)+\log|{{WW}}^{T}+\zeta^{2}{I}|+\text{tr}\{({WW}^{T}+\zeta^{2}I)^{-1}\hat{\Phi}\}],\label{eq:loglik}
\end{equation}
where $\hat{\Phi} = m^{-1}\sum_{j=1}^m(X_j - \hat{\mu})(X_j - \hat{\mu})^{T}$ is the sample covariance matrix. Assuming $m > n$ and $\hat{\Phi}$ is full rank, the maximum likelihood estimator (MLE) for $\mu$ is simply the sample mean $\hat{\mu} = \frac{1}{m}\sum_{j=1}^m X_j$. Without loss of generality, $\hat{\mu}$ can be replaced by zero provided that the data $X$ had been row centred. For convenience, we also assume the data had also been row standardized such that the diagonal elements of $\hat{\Phi}$ equal to 1.

This intrinsic data dimension, $\operatorname{rank}(W) = \operatorname{rank}(W^{T}W)=k$, is only implicitly involved in the log-likelihood. It has been shown in~\citep{Tipping1999a} that for any integer $q \in \{1,\ldots,n-1\}$, \eqref{eq:loglik} is maximized by:
\begin{equation}
\hat{W}_q =U_q\hat{D}{_q} B{_q}, \quad \text{and} \quad \hat{\zeta}^{2}_q=\frac{\sum^{n}_{i=q+1}{\hat{\lambda}_{i}}}{n-q}, \label{eq:ppca_mle}
\end{equation}
where $\{\hat{\lambda}_{i}\}_i$'s are the sample eigenvalues of $\hat{\Phi}$, $U_q$ is an $n \times q$ matrix with columns corresponding to the first $q$ eigenvectors of $\hat{\Phi}$, $\hat{H}(q)$ is a diagonal matrix with the first $q$ non-zero entries each given by $\hat{\eta}_{i} = \sqrt{{\hat{\lambda}_{i}} - \hat{\zeta}^{2}_q}$, and $B_q \in \mathbb{R}^{q\times q}$ is an arbitrary orthogonal matrix. The integer $q$ needs not be specified, but the form of~\eqref{eq:ppca_mle} suggests that the division between the first $q$ and the last $n-q$ eigenvalues/eigenvector is the key to maximizing~\eqref{eq:loglik}. In other words, for every value of $q$, we can identify the corresponding MLEs given in~\eqref{eq:ppca_mle}, but the different choices of $q$ cannot be distinguished under the current likelihood model.
 
Let $l_p$ denote the profile log-likelihood. If we considered the parameters $W, \sigma^2$ to be nuisance parameters, a profile log-likelihood in $q$ is obtained by substituting the solutions in~\eqref{eq:loglik}:
\begin{equation}
l_p(q; \hat{\lambda}_i) = -\frac{m}{2}\{n\log(2\pi)+\sum_{i=1}^{q}\log{\hat{\lambda}_{i}}+(n-q)\log\hat{\zeta}_q^{2}+n\}. \label{eq:maxlk}
\end{equation}

The formulation~\eqref{eq:maxlk} clarifies that data dimension is implicitly involved in defining the parameters of the model, and one might be tempted to find the maximizer (in $q$) of the profile log-likelihood as the estimate of effective rank. However, the following result suggests that the profile log-likelihood alone is not sufficient to identify the intrinsic data dimension.

\begin{proposition}
\label{Proposition1}
Consider a sample $X \in \RR^{n \times m}$ with each column following a multivariate Gaussian distribution $N(0, {W}{W}^{T} + \zeta^2{I})$. If the sample row covariance matrix of $X$ is positive semi-definite and $k = \operatorname{rank}({WW}^{T})$, then the profile log-likelihood $l_{\text{p}}(q)$ is non-decreasing in $q \in \mathrm{Z}_+ (1 \le q \le n-1)$.
\end{proposition}

Proof is included in Supplementary Materials.

This result shows that the profile log-likelihood is monotonically non-decreasing in $q$, suggesting that it can not be used as a criterion to select $k$, the data dimension, in finite samples. The choice of $k$ thus becomes a model selection-type problem, with decreasing values of $q$ corresponding to more constraint models and $q=n$ corresponds to a fully non-parametric, conventional PCA.

\begin{remark}\label{remark:p1}
Proposition~\ref{Proposition1} demonstrates that the saturated model with $q = n-1$ is always preferred. If one permits $q = 0$, then $\hat{W} = \mathbf{0}$ and the likelihood is minimized. The same conclusion can be reached by observing the proportion of variance explained by the PPCA model with true rank $k$:
\begin{equation*}
\text{tr}(WW^{T}) = {\sum_{i=1}^{k} d_i^2} =  n(1- \zeta^2),
\end{equation*}
where $\{d_i\}_{i=1, \dots, k}$ are the singular values of $W$. When $\zeta^2$ is equal to 0 (or $k = n$), the model corresponds to PCA with a full-rank loading matrix and is completely deterministic; and when $\zeta^2$ is equal to 1 (or $k=0$), the model reduces to an isotropic Gaussian distribution and $W = \mathbf{0}$. In order to avoid degenerate situations, in this paper we restrict the range of $k$ to $\{1,2,\ldots, n-1\}$.
\end{remark}

\begin{remark}\label{remark:p2}
The generative model~\eqref{eq:generative_model} has a specific dimension $k$, which is embedded in the parameter $W$ through the data generative process. At the same time, the data generated can support each possible $q$ if we evaluate the model likelihood alone without any constraint on the error variance or model complexity.
\end{remark}


\section{Effective rank selection heuristics based on a penalized probabilistic principal components analysis} \label{s:pppca}

Penalized maximum likelihood approaches are widely used to induce sparsity in statistical models. The level of penalty imposed on the model is regularized via a tuning parameter, which controls the trade-off between goodness-of-fit and complexity~\citep{10.2307/2346178, Zou2006sparse, Bien2011}. In the problem considered here, the model complexity, defined by the number of free parameters $nk + 1 - k(k-1)/2$, is directly related to the data dimension, while the fit corresponds to the amount of variance explained, i.e.\ $\text{tr}(\Phi) - n\zeta^2$. The natural guiding principle is to favour a parsimonious representation for the covariance by simultaneously penalizing small explained variance and large $k$.

The penalized log-likelihood has the form:
\begin{equation*}
l(W, \zeta^{2}; \delta) = \frac{-m}{2}\{\log|{WW}^{T}+\zeta^{2}{I}|+\text{tr}[({WW}^{T}+\zeta^{2}{I})^{-1}\hat{\Phi}]-\delta \text{pen}(W, \zeta^2) \}, 
\end{equation*}
where the tuning parameter $\delta>0$ controls the amount of penalty due to a penalty function, $\text{pen}(W, \zeta^2)$. Notice that $m$ is a scaling factor and does not directly affect the maximization other than through the convergence of $\hat{\Phi}$ to the true covariance $\Phi$.

The penalty function should depend on $(W, \zeta^2)$ and thus be able to capture the model dimension embedded in $W$ and the amount of error variance $\zeta^2$. At the same time, the two parameters combine in the case of standardized data because $\text{tr}(WW^{T}) + n\zeta^2 = n$. By maximizing the penalized log-likelihood function, it will also be possible to express the penalized MLEs indexed by $q$ and thus to motivate the penalized profile log-likelihood as a vehicle for intrinsic data dimension selection.

Unlike in other constrained optimization problems, the estimation of individual entries of $W$ is not the primary objective. Rather, we are interested in penalty functions that diverge when the estimated eigenvalues (i.e.\ the sum of $\zeta^2$ and each squared singular value of $W$) are close to 1, or alternatively, when $\zeta^2$ is close to 0. Here we explore the following penalty functions that capture both the amount of variance explained and the complexity of the model:
\begin{subequations}
\begin{align}
 \text{pen}_1(W, \zeta^2) &= \operatorname{rank}({W}) \log{\zeta^2} = k\log{\zeta^2} \label{eqn:pen} \\
 \text{pen}_2(W, \zeta^2) &= -\frac{\operatorname{rank}(W)}{\zeta^2} = -\frac{k}{{\zeta^2}}  \label{eqn:pen2}\\
 \text{pen}_3(W, \zeta^2) &= \beta  \text{pen}_1(W, \zeta^2) + (1-\beta)\text{pen}_2(W, \zeta^2), \quad \beta \in (0,1).  \label{eqn:pen3}
 \end{align}
\end{subequations}

In our experience, the penalties lead to equivalent analyses since the tuning parameters will adjust to yield similar results. Ultimately, the choice to use~\eqref{eqn:pen} over the others is driven by convenience because it leads to simpler analytical derivations and intuitive heuristics.

\subsection{Penalized maximum likelihood estimators}

The penalized log-likelihood using the proposed penalty function~\eqref{eqn:pen} becomes:
\begin{align}
\nonumber l(W, \zeta^{2}; \delta) & =\frac{-m}{2}\{\log|{WW}^{T}+\zeta^{2}{I}|+\text{tr}[({WW}^{T}+\zeta^{2}{I})^{-1}\hat{\Phi} ]-\delta \operatorname{rank}(W)\log{\zeta^2} \}. 
\end{align}

Similarly to~\eqref{eq:ppca_mle}, the penalized MLEs, $\tilde{W}$ and $\tilde{\zeta}^2$, are functions of $q$. Due to a non-zero $\delta$-value, the penalized MLE of $\zeta^2$ is expressed in terms of $\delta$ and $\hat{\zeta}^2_q$:

\begin{equation}
\tilde{\zeta}^2_q =  \frac{\sum_{i=q+1}^n \hat{\lambda}_i}{n-q-\delta q} =  \frac{n-q}{n-q-\delta q}\hat{\zeta}^2_q.\label{eq:sigmak}
\end{equation}

Taking derivative with respect to $W$ yields the same relationship between the squared singular values of $W$ and $\zeta^2$:
\begin{equation}
   \hat{\lambda}_i=
\begin{cases}
    \tilde{\eta}^2_i(q)+\tilde{\zeta}^2_q,  & \text{if } i \le q; \\
    \tilde{\zeta}^2_q,              & \text{otherwise},
\end{cases}\nonumber
\end{equation}
where $\tilde{\eta}^2_i(q)$ denotes the $i$th estimated value when the estimated effective rank is $q$. For a fixed $q$, $\tilde{\zeta}^2_q$ is unbounded as $n-q-\delta q$ can be very close to 0 or even negative for large $\delta$-values. This implies that the choice of $q$ poses a restriction of the range of $\delta$, and vice versa. Thus, the theoretical range of $\delta$ has an upper bound at $n/q-1$ so that $\tilde{\zeta}^2_q$ is positive. Henceforth, we reparametrized the tuning parameter to $\tilde{\delta} = \delta/n \in [0,1/q-1/n)$.

Interestingly, the penalized MLEs of $\zeta^2$ under \eqref{eqn:pen} and~\eqref{eqn:pen2} are closely related to those estimated under an approximated posterior likelihood assuming an inverse-gamma prior~\cite{minka2001automatic}, with $\tilde{\delta}$ corresponding to linear functions of the hyperparameters, see Appendix~\ref{ss:pppca_minka} for more details. 

Substituting the penalized MLEs given $q$, we obtain the penalized profile log-likelihood, denoted by $l_p(q;\tilde{\delta})$, as a function of $q$ for a fixed $\tilde{\delta}$:
\begin{equation}
l_p(q;\tilde{\delta}) = l_p(q) -\frac{m}{2}\left[n\Big(1-\frac{q}{n}-\tilde{\delta} q\Big)\log{\frac{n-q}{n-q-nq\tilde{\delta}}} -\tilde{\delta}  nq(\log{\hat{\zeta}^{2}_q}+1)\right].
\label{eq:penloglik}
\end{equation}

The penalized profile log-likelihood criterion favours a more parsimonious model by penalizing large values of $q$ as well as retaining as much explained variance as possible. Given suitable choices of $\tilde\delta$, the following results establish the conditions under which the penalized criterion will find the correct dimension:

\begin{proposition}\label{Proposition2}
Consider a sample $X \in \mathbb{R}^{n \times m}$ with each column following a multivariate Gaussian distribution $\mathcal{N}(0, WW^{T} + \zeta^2{I})$. If $\hat{\Phi}$, the sample covariance matrix of $X^{T}$, is positive semi-definite, then there exists $\tilde{\delta}_o \in (0, 1-1/n)$ such that $l_p(q;\tilde{\delta}_o)$ is maximized at $k$ $(1 < k < n)$, the rank of $W$ or the effective rank of $X$.
\end{proposition}

Proof is included in Supplementary Materials.

\subsection{A data-driven voting strategy to estimate the effective rank}

The introduction of penalty changes the monotonicity property of the profile likelihood~\eqref{eq:maxlk}, and thus makes it possible to select the correct dimension $k$ for appropriate choices of $\tilde{\delta}$-value. The selection of appropriate tuning parameter values in other well-known problems, such as the selection of shrinkage tuning parameter in lasso~\citep{10.2307/2346178,RSSB:RSSB693}, uses either a model selection criterion, e.g.\ Akaike or Bayesian information criterion, or cross-validation. However, the use of a cross-validation approach is based on optimizing a certain objective function that can be analytically expressed, a task that is difficult when of interest is determining the dimension. Our attempts at using an off-the-shelf information criterion produced modest results in simulations under the correct model specification, but failed to identify a sensible estimate when the data generative model deviated from assumptions.

So far, a data-driven heuristic gave the best balance in performance. It entails a voting strategy in which each value of $\tilde{\delta}$ over a plausible range, determined from the data, will lead to a vote for a particular value of $q$ as the estimate. Since the same estimate of $k$ can result from multiple $\tilde{\delta}$-values, ultimately the estimated dimension that has been obtained most often is selected. 

The search for the intrinsic dimension implies a grid search for $\tilde{\delta}$ whose values $\{\tilde{\delta}_{1},\ldots, \tilde{\delta}_{T}\}$ are selected using a sequence of $T$ equidistant points on $\log$ scale. The user-specified integer $T$ needs to be large enough to identify a mode, and in simulations we used $T=5,000$ or roughly $50n$, with values of the same order of magnitude leading to the same results. Each $\tilde{\delta}_t$ will result in~\eqref{eq:penloglik} supporting a possible value for $k$ $(1 \le k \le n-1)$ , which is the maximizer of $l_p(q; \tilde{\delta}_t)$ in $q$. Then, the number of times that a value of $k$ maximizes the penalized profile log-likelihood is counted and the one with the highest vote count is selected. Define $|A(j)| = \#\{t: \operatorname{arg\,max}_q l_p(q,\tilde\delta_t)=j\}$ and the estimate is denoted by $\tilde k = \operatorname{arg\,max}_j |A(j)|$. The data-driven procedure is described in Algorithm~\ref{alg:wordy}.

\begin{algorithm}[hbt!]
\caption{A data-driven voting strategy to estimate the effective rank}\label{alg:wordy}
\begin{algorithmic}
\Require{integer $T$, $\{\hat{\lambda}_i\}_{i=1,\dots, n-1}$, $\kappa = 0.001$}
 \State initialization; setting $n_{\max} = \min \{i: \hat{\lambda}_i < \kappa\}-1$
\If{$n_{\max} > 1$}
 \While{$q = 1$}
  \State find $\tilde{\delta}_{T} = \operatorname{arg\,min}_{\tilde{\delta}} \{u: l_p(q=1;u) > l_p(q=2;u)\}$;
    \EndWhile
  \While{$q = n-2$}
  \State find $\tilde{\delta}_{1} = \operatorname{arg\,max}_{\tilde{\delta}} \{u: l_p(q=n-2;u) > l_p(q=n-1;u) \}$;
      \EndWhile
   \State construct $\{\tilde{\delta}_{1},\ldots, \tilde{\delta}_{T}\}$;
   \While{$j \le n-1$}
   \State $|A(j)| = \#\{t: \operatorname{arg\,max}_q l_p(q,\tilde{\delta}_{t})=j\}$
     \EndWhile
      \State $\tilde{k} = \operatorname{arg\,max}_j |A(j)|$
  \Else
   \State $\tilde{k} = 1$
  \EndIf
\end{algorithmic}
\end{algorithm}

The penalized approach requires a proper calibration of $\tilde{\delta}$ so that the true dimension, $k$, identifies as the global maximizer of $l_p(q;\tilde{\delta})$ most often. In theory, $\tilde{\delta}$ could take any value in $[0, \infty)$, but for practical considerations, it has a finite range depending on the maximum and minimum $q$ to avoid degenerate cases. The connection between $q$ and $\tilde{\delta}$, given by $\tilde{\delta} \in (0,(1/q-1/n)[1-\hat{\zeta}^{2}_q])$, is derived in Appendix~\ref{ss:pPPCA_details}. A theoretical justification of the voting method based on the $\log$-scale is provided in Lemma~\ref{lemma_25} in Appendix~\ref{ss:pPPCA_details}. A detailed illustration of the method on simulated data can be found in Supplementary Materials. 

To make the methods accessible, we implemented the voting procedure in a statistical software \texttt{R} package, available at \url{https://github.com/WeiAkaneDeng/SPAC2}.

\section{Simulation studies}\label{s:sim_res}

\subsection{Data simulation}

Given the true dimension $k$, error variance $\zeta^2$, and observed dimensions $(n, m)$, we can generate the data by specifying either 1) the signal components of the first $k$ true eigenvalues ($\eta^2_1, \dots, \eta^2_{k}$) directly or, 2) a trend in the first $k$ signal components. The residual noise was assumed to have a multivariate distribution with mean vector zero and covariance $\zeta^2{I_n}$. The maximum dimension ($n$, when $n < m$) is often directly associated with the difficulty of recovering the true dimension and was kept fixed at $n=100$. 

We explored four data generation scenarios: the first scenario, denoted by S0, is a baseline case where each observation is independent and identically distributed (i.i.d) following a standard normal distribution; the second scenario encompassed the spiked covariance model with either the first $k$ true eigenvalues being equal, a homogeneous setting (scenario S1.1), or decaying with a linear or an exponential trend, the heterogeneous settings (scenario S1.2); the third scenario, S2, explored varying data dimensions whereby the row covariance matrix could also be rank-deficient; and finally, scenario S3, examined the impact of model violations, such as heavy tails and correlated observations. These scenarios are summarized in Table~\ref{table:sim}. For each condition, the simulation was repeated 100 times and the number of observations was fixed at $m=5,000$ except in scenario S2. Though there is no explicit assumption requiring $m>n$, the choice for a larger $m$ is to ensure some consistency in the sample eigenvalues, which is essential to the majority of the methods.

We applied double standardization to each simulated dataset and then calculated the sample eigenvalues. For data generated under S3, the sum of the sample eigenvalues could potential exceed $n$ as data deviated from normality, thus the sample eigenvalues were scaled to sum to $n$ prior to analysis. Meanwhile, when the row covariance is rank-deficient, the trailing sample eigenvalues could be negative; in this case, we adjusted the search space to $\{1, 2, \dots, n_\text{max}\}$, where $n_\text{max} = \max_{i}(\hat{\lambda}_i > 0.001)$. 

\begin{table}[H]
\begin{tabular}{llll}
\toprule
\textbf{Scenario}            & \textbf{Description of scenarios}                                                     & \textbf{Data dimensions}  & \textbf{Error distribution} \\
\hline
S0                  & i.i.d.  & $n < m$    & $\mathcal{N}(0,1)$             \\
\hline
\multirow{2}{*}{S1} & Homogeneity                                                                  & $n < m$   & $\mathcal{N}(0,\zeta^2)$             \\
                    & \begin{tabular}[c]{@{}l@{}}Heterogeneity\\ (linear/exponential)\end{tabular} & $n < m$   & $\mathcal{N}(0,\zeta^2)$              \\
\hline
S2                    & \begin{tabular}[c]{@{}l@{}}Heterogeneity \\ (exponential)\end{tabular}       &$n > m$ and $n < m$ & $\mathcal{N}(0,\zeta^2)$             \\
\hline
\multirow{3}{*}{S3} & heavy tails & $n < m$    & $t(\text{df})$, $\text{df}=\{3,5,10\}$             \\
                    & correlated observations                                                      & $n < m$   &  $\text{AR1}(\rho)$, $\rho = \{0.1,0.4, 0.7\}$              \\
                    & both                                                                         & $n < m$    &   $t(\text{df})$ and $\text{AR1}(\rho)$     \\
\bottomrule
\end{tabular}
\caption{Simulation scenarios.} \label{table:sim}
\end{table}

\subsection{Alternative methods}

The performance of the proposed approach, denoted by \textit{pPPCA} for penalty~\eqref{eqn:pen}, is compared with a list of alternative methods (mathematical constructions in Appendix~\ref{append:methods}). For completeness, we also included \textit{pPPCA2} for penalty~\eqref{eqn:pen2}, and \textit{pPPCA3} for penalty~\eqref{eqn:pen3} with $\beta = 1/2$. Briefly, we focused on the class of model selection criteria, including Akaike information criterion (\textit{AIC}); a simplification to the Laplace's method using BIC approximation \citep{kass1995bayes}, denoted by \textit{BIC}; an approximation to the posterior likelihood using Laplace's method proposed in \cite{minka2001automatic}, denoted by \textit{Laplace}; the best performer from a class of Bayesian criteria under different diverging assumptions, PEnalized Semi-integrated Likelihood (\textit{PESEL}; \citealp{sobczyk2017bayesian}). There is another class of methods that focused on the estimation, including a bias-corrected criterion for estimating $k$ by~\cite{passemier2017estimation}, denoted by \textit{Passemier}, and a list of Bai and Ng's criteria~\cite{Bai2002}, denoted by \textit{BN}. A hypothesis testing criterion for the equality of the last $n-k$ eigenvalues \cite{lawley1956tests} was also selected, denoted by \textit{Lawley}. The hard threshold-based approach~\cite{gavish2014optimal} removes the underlying noise in the singular values, and is denoted by \textit{Donoho}. Finally, the last class of methods attempt to detect an ``elbow'' in the scree plot produced by the sample eigenvalues: a list of empirical approaches, as well as a simple profile likelihood-based criterion (\textit{ProfileL}) by~\cite{zhu2006automatic} were included in the comparison.

Berthet and Rigollet~\cite{berthet2013optimal} considered the minimal value of $\theta (>0)$ in a more restrictive spiked covariance model $I + \theta vv^{T}$ that can be theoretically distinguished from $I$, where $v = (v_1, \dots, v_k)$ is a set of $n$-dimensional unit vectors. This is equivalent to our problem when the top $k$ eigenvalues are equal. For each true $k$, a corresponding critical value is given and shown to be of order $k\sqrt{log(n/k)/m}$~\citep{berthet2013optimal}, implying that as the true $k$ increases, the signal needs to increase relatively for detection. Results from this study, though not directly applicable for method comparison, provide insight for the simulation study that follows.

Some of the methods we do not consider in the comparison are automatic relevance determination~\citep{Bishop19991} and related methods that followed it \mbox{\citep{everson2000inferring,rajan1997model}} as they have been shown to be outperformed by methods based on the Laplace approximation \citep{minka2001automatic}. Variational approximation methods~\citep{Bishop19992,ilin2010practical, nakajima2012perfect} are also excluded, as \cite{ilin2010practical} does not directly estimate the number of PCs, while \cite{nakajima2012perfect} has been shown to be suboptimal to \cite{hoyle2008automatic}. We have also excluded Bayesian methods that rely on MCMC sampling \citep{hoff2007model}, as they become computationally prohibitive when either $n$ or $m$ is large ($>1,000$). The large number of observations is why cross-validation is difficult to implement beyond the heavy computational burden as data splitting can sometimes create biased signal in the data depending on how the held-out datasets are obtained, i.e.\ when the covariance structure is local to a subset of the observations. For this reason, we excluded cross-validation, but included the general cross-validation (\textit{GCV}) criterion of \cite{josse2012selecting} that has better scalability properties.

\subsection{Scenario 0: Independent identically distributed}

\begin{figure}
\centering
\includegraphics[scale=0.125]{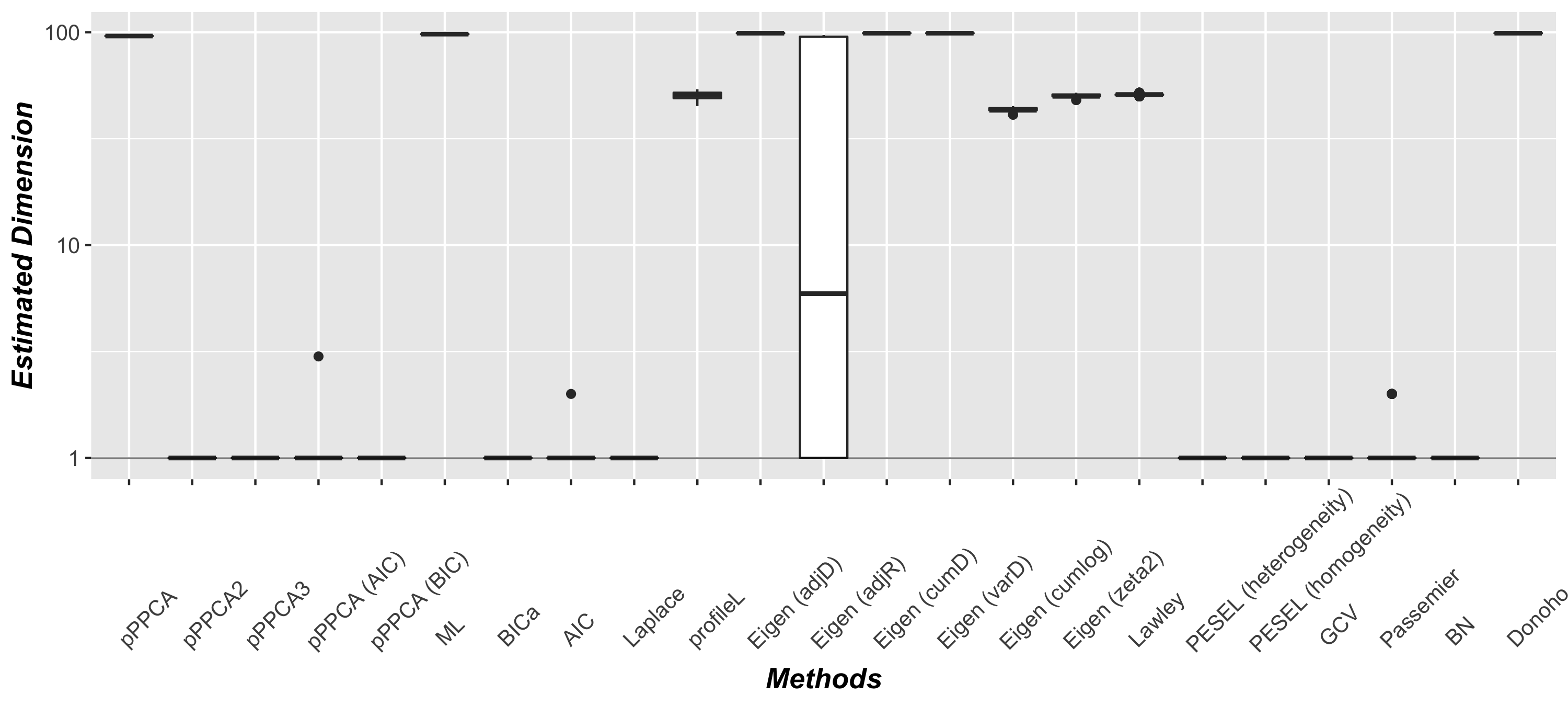}
\caption{Distribution of the estimated $k$ over 100 replicates when data are i.i.d.} \label{fig:S0}
\end{figure}

As a baseline scenario, we compared methods when the data were drawn from a multivariate normal distribution with zero mean and an identity covariance. Depending on what is considered independent signal and noise, the effective rank could be 0 or a value close to the maximum possible rank $n-1$ (due to the standardization). Unsurprisingly, most methods estimated either 1 or $n-1$ majority of the time (Figure~\ref{fig:S0}), with \textit{pPPCA} preferring $n-1$ and most other model selection methods choosing 1. In this case, \textit{Lawley}, \textit{profileL}, and some ``elbow''-based empirical approaches do not work very well, giving estimates ranging between $60-80$, capturing the fluctuation in sampling distribution of the bottom eigenvalues.

\subsection{Scenario 1.1: Homogeneous eigenvalues}

The first experiment consisted of $k$ equal squared singular values, where we used $\zeta^2_{k} = \{0.8, 0.81, \dots, 0.99\}$ and $k= \{5, 10, 20\}$ to capture a range of signal to noise (SNR) values, defined by the ratio of $\eta^2_{k}$ and $\zeta^2_{k}$ rather than $(1-\zeta^2_{k})\zeta^{-2}_{k}$. The theoretical lower bounds of $k\sqrt{log(n/k)/m}$ roughly correspond to $\zeta^2_{k}=0.98$ for $k = 10$ and $\zeta^2_{k}=0.93$ for $k = 20$. 

The best performer from each class of methods is presented in Figure~\ref{fig:S11_sub}. The results of all methods can be found in Supplementary Figure 1. Most methods exhibited a decreasing relationship between correctly estimated  dimension as a function of increasing SNR (i.e.\ small $\zeta^2$ and small $k$), with the exception of \textit{AIC} and \textit{Passemier}, where both methods have good performance most of the time. Interestingly, though \textit{pPPCA} showed a decreasing trend as SNR increased for each true $k$, its performance did not deteriorate terribly across the different true $k$. The other methods were clearly more sensitive to the SNR as they approached the theoretical lower bounds for detection, with \textit{AIC}, \textit{Laplace} having the best performance among model selection approaches and \textit{profileL} and \textit{Eigen (adjR)} having the best performance among empirical approaches. The two \textit{PESEL} criteria were similar to \textit{BIC} and both had better performance than \textit{pPPCA}. 

We expected methods that take advantage of the homogeneity in the true eigenvalues to have reasonably good performance, such as \textit{profileL} and \textit{PESEL (homogeneity)}. But in fact, \textit{profileL} was better than \textit{PESEL (homogeneity)} as SNR approached the theoretical minimum at $\zeta^2_{k}=0.98$ for $k = 10$, and even better than \textit{Laplace} when $k = 20$.
 
At this point, we can eliminate both \textit{Donoho} and \textit{ML} from the list of methods as they were not designed to detect the effective rank, as the former aims to detect a theoretical minimum in terms of mean squared error loss, while the latter is a flawed information measure for model selection. 

\begin{figure}
\centering
\includegraphics[scale=0.125]{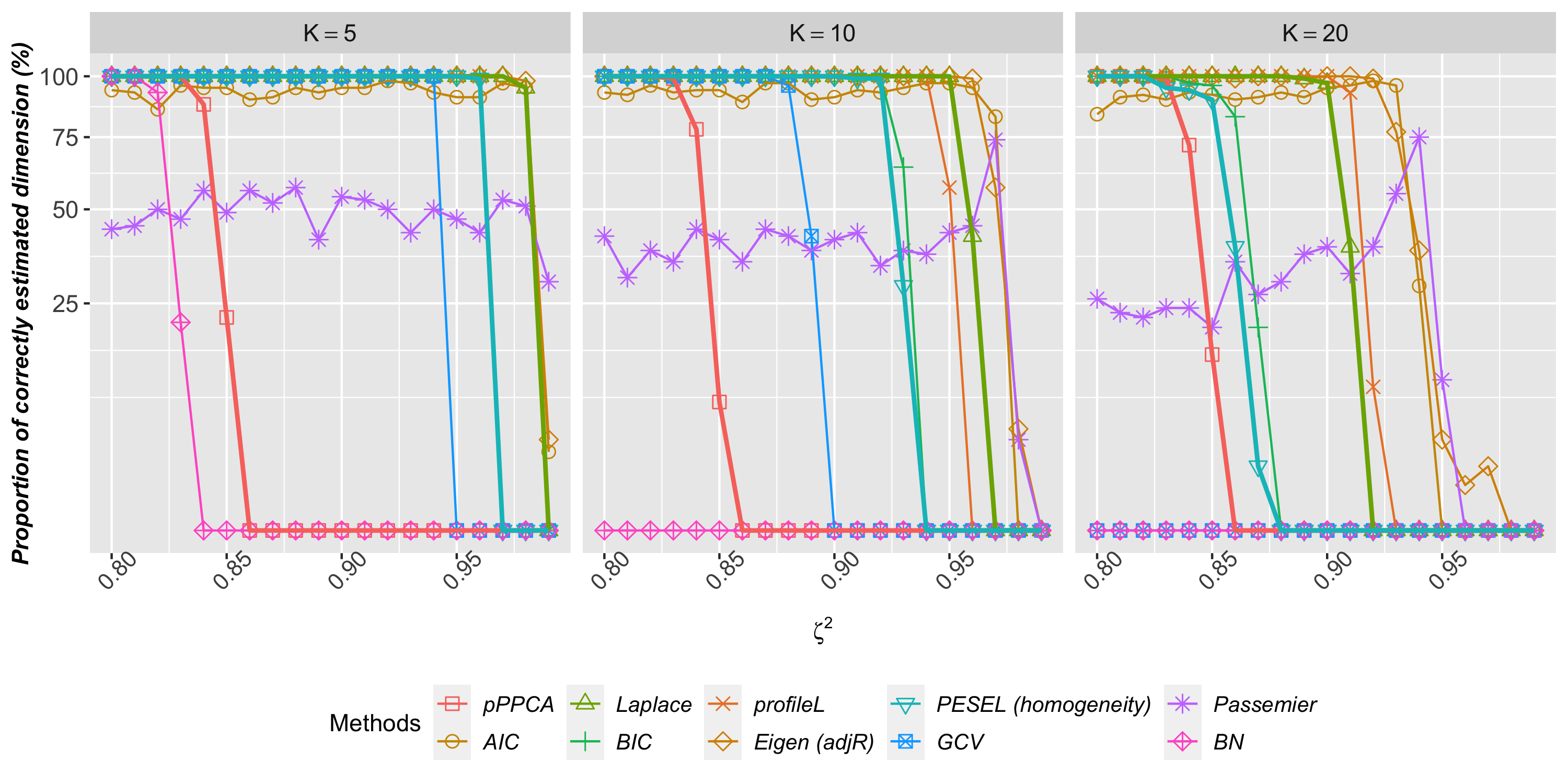}
\caption{Proportion of correctly estimated $k$ over 100 replicates as a function of $\zeta^2$ assuming the first $k$ squared singular values are equal. The colored line corresponds to each method among the subset with reasonably good performance.} \label{fig:S11_sub}
\end{figure}

\subsection{Scenario 1.2: Heterogeneous eigenvalues}\label{ss:s12_het}

A more interesting and realistic scenario is when the true eigenvalues decrease according to a linear or exponential trend. In this case, the singular values can be determined by varying the two parameters $\zeta^2_{k}$ and $\eta^2_{k}$ for a given $k$. We chose $\zeta^2_{k} = \{0.1, \dots, 0.8\}$, $\eta^2_{k} = \{0.1, 0.3\}$, and $k = \{5, 10\}$. The performance of methods could possibly be impacted by the following factors, including 1) the trend in $\{ \eta_{i}^2\}_{i=1, \dots, k}$, the signal components, 2) true dimension $k$, and 3) the error variance $\zeta^2$.

\begin{figure}[H]
\centering
\includegraphics[scale=0.12]{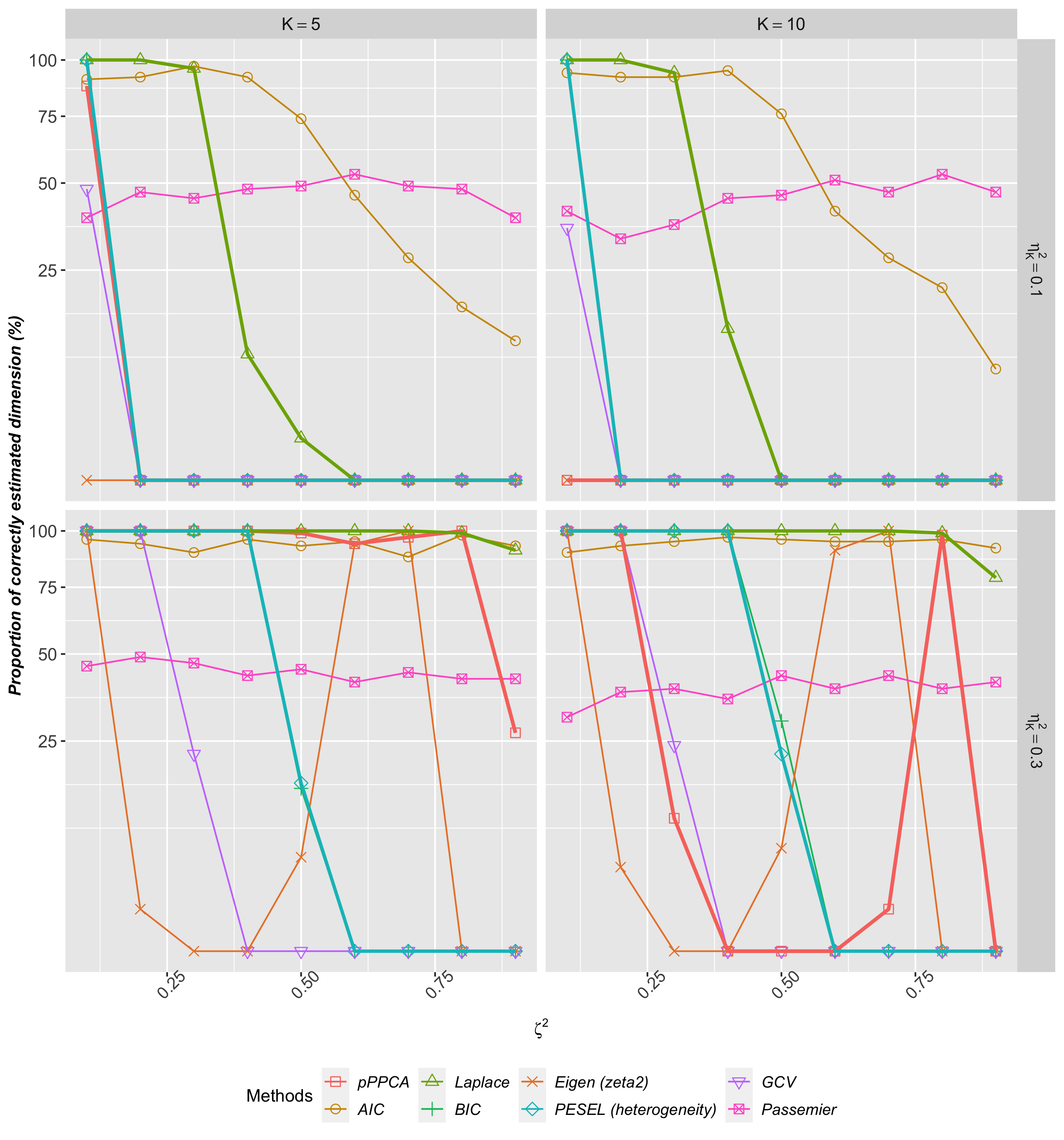}
\caption{Proportion of correctly estimated $k$ over 100 replicates as a function of $\zeta^2$ assuming a linear decay in the first $k$ squared singular values.}\label{fig:S12_linear_sub}
\end{figure}

\begin{figure}[H]
\centering
\includegraphics[scale=0.12]{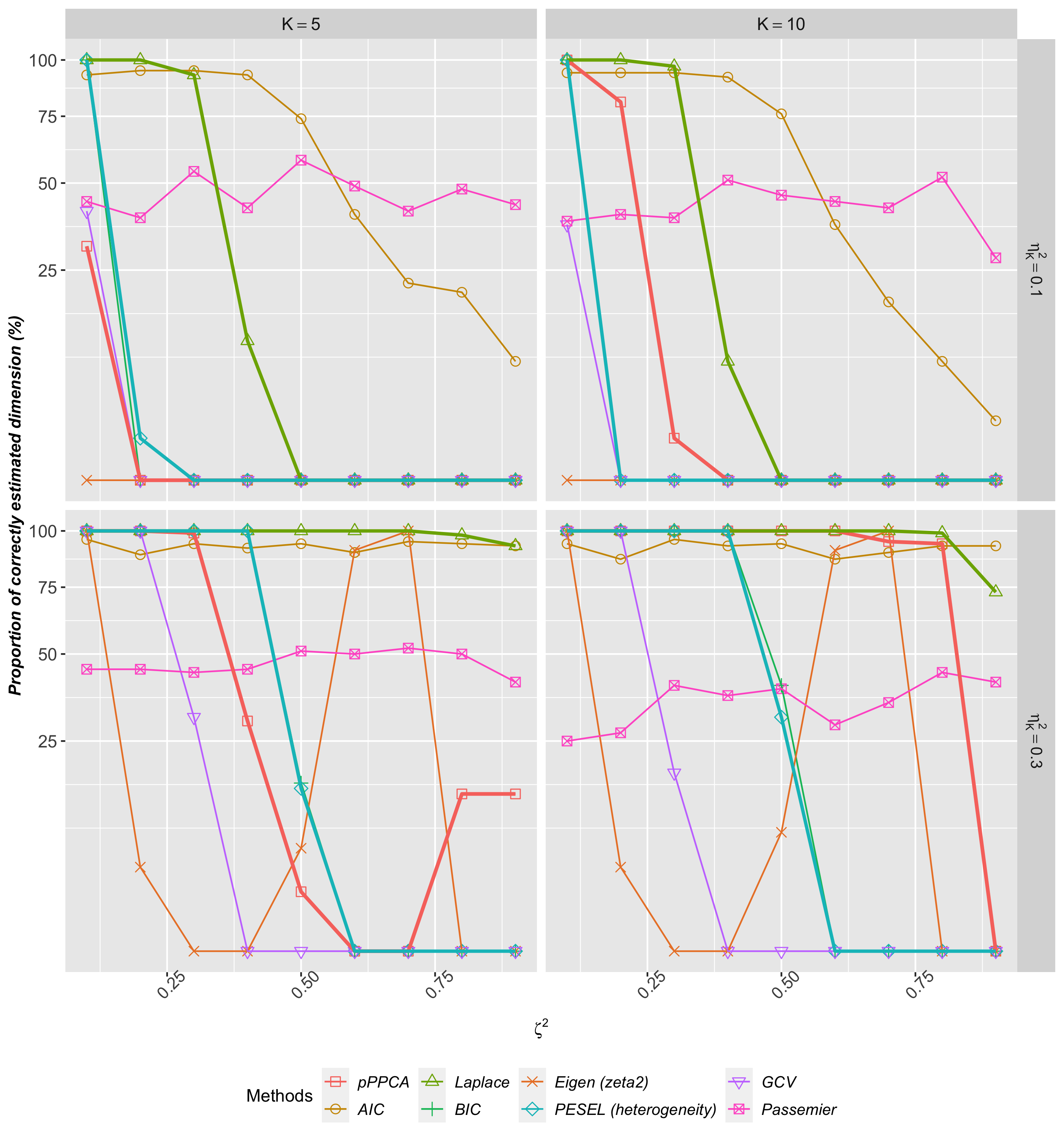}
\caption{Proportion of correctly estimated $k$ over 100 replicates as a function of $\zeta^2$ assuming an exponential decay in the first $k$ squared singular values.}\label{fig:S12_exp_sub}
\end{figure}

\textit{Laplace} had the best performance across the conditions, followed by the proposed \textit{pPPCA}, \textit{PESEL (heterogeneity)}, where both method would underestimate by 1. For most methods, we observed little impact on the performance of methods due to the choice of a linear and an exponential trends (Figures~\ref{fig:S12_linear_sub} and~\ref{fig:S12_exp_sub}). However, performance of \textit{pPPCA} was superior for a linear trend when true $k=5$ (Figures~\ref{fig:S12_linear_sub}) or an exponential trend (Figure~\ref{fig:S12_exp_sub}) for a larger $k = 10$, possibly related to the fact that the empirical range of the penalty parameter influenced the sampling distribution of the first $k$ sample eigenvalues.

Contrary to the homogeneous case, the decreasing trend in the signal component posed difficulty most noticeably for methods that assumed homogeneity. For example, both \textit{profileL} and \textit{PESEL (homogeneity)} completely failed to recover the correct dimension and underestimated. Again, we observed \textit{PESEL (heterogeneity)} to be near identical to \textit{BIC} and that \textit{AIC} and \textit{Passemier} would estimated correctly most of the time, but both are inconsistent.

\subsection{Scenario 2: Data dimensions}

One of the data attributes encountered in real world applications is the varying ratios of $m$, the number of observations, and $n$, the maximum dimension. To evaluate the performance with respect to different ratios, we assumed the first $k$ ($=\{5, 10\}$) squared singular values were equal (i.e.\ homogeneous) or decayed linearly or at an exponential rate with their values determined by fixing $\eta^2_{k} = 0.3$, $\zeta^2_{k} = 0.5$. The choice of $m$ was set to be 50, 500, 1,000, 5,000, 10,000, and 20,000. 

\begin{figure}[H]
\centering
\includegraphics[scale=0.125]{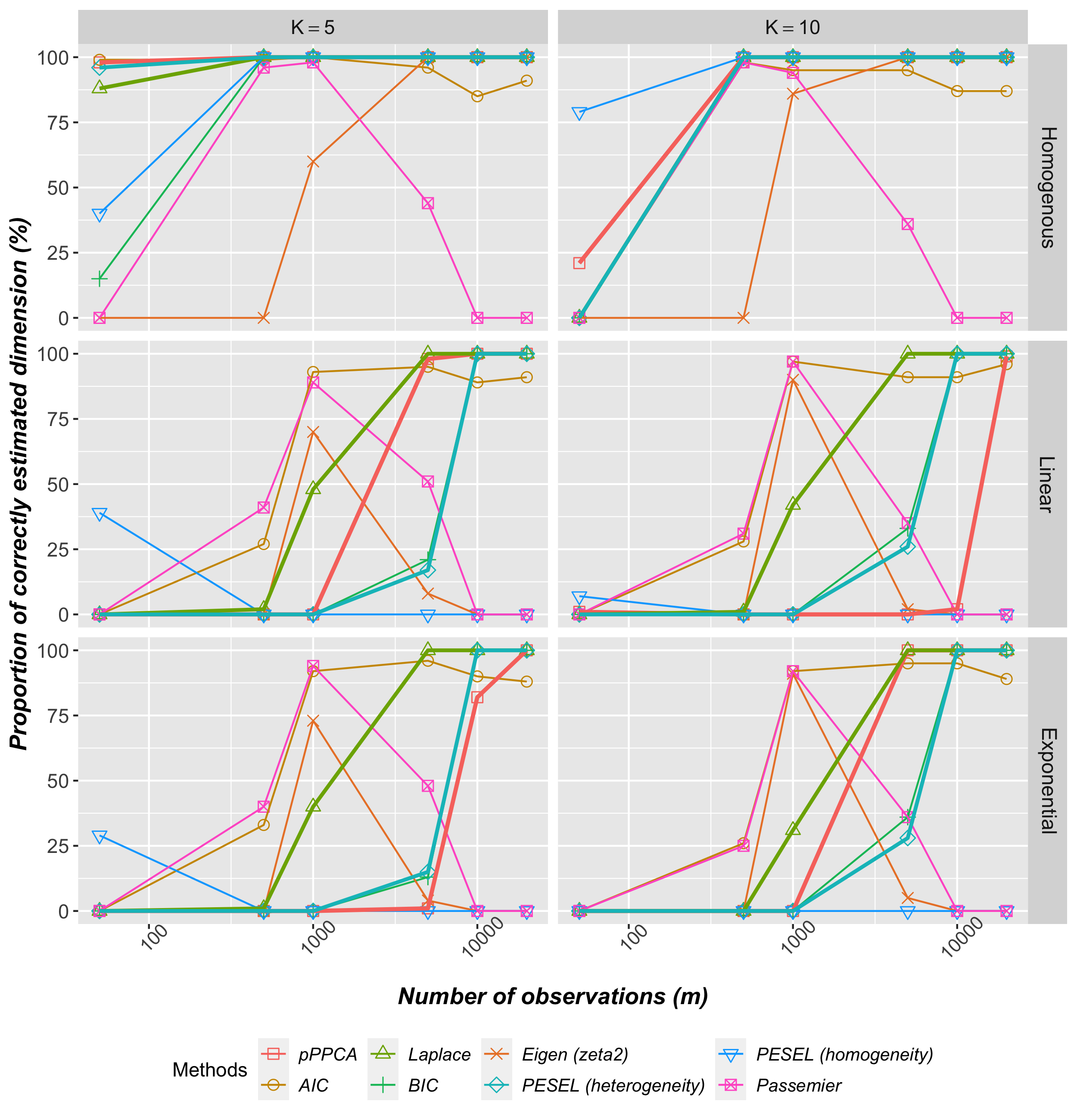}
\caption{Proportion of correctly estimated dimension over 100 replicates as a function of $m$ assuming homogeneity, a linear or an exponential decay in the first $k$ squared singular values.} \label{fig:S2}
\end{figure}

Informed by results in Section~\ref{ss:s12_het}, we compared only methods that correctly estimated at least 5\% for this slightly challenging scenario, including \textit{AIC}, \textit{BIC}, \textit{Eigen ($\zeta^2$)}, \textit{Laplace}, \textit{Passemier}, \textit{PESEL (heterogeneity)}, and \textit{pPPCA}.

As $m$ was increased, estimates from \textit{BIC}, \textit{Laplace}, \textit{PESEL (heterogeneity)}, and \textit{pPPCA} all approached the correct dimension 100\% (Figure~\ref{fig:S2}). Across different $m/n$ ratios, \textit{pPPCA} had the best performance when the signal was homogeneous; while there was no dominant method when the signals were heterogeneous, \textit{Passemier}, \textit{AIC} or \textit{Laplace} were competitive depending on values of $m/n$. Among methods that are empirically consistent, \textit{Laplace} had superior performance than both \textit{PESEL (heterogeneity)} and \textit{pPPCA}. Between these two, there was no universally better method across the combinations of $k$ and linear/exponential trends. Again, we observed that the type of trend has a bigger impact on the performance of the proposed \textit{pPPCA} than other methods, preferring a linear trend when $k=5$ or an exponential trend for a larger $k = 10$.

\subsection{Scenario 3: Departure from model assumptions}

In many applications, noise in the data might not be independently or normally distributed. We investigated cases where the observed error was drawn from a \textit{t}-distribution and with correlation as typically observed in genetic data where the strength of correlation between pairs of genetic features increases with proximity. Since the features are order-invariant, we simulated from a covariance with a block structure driven by an auto-correlation process.

Let $\nu$ denote the degrees of freedom for the \textit{t}-distribution and $\rho$ denote the auto-correlation coefficient. The error $f_j$, for $j = 1, 2, \dots, m$, was generated according to
\begin{equation*}
f_j = f_{j-1}\rho + r,
\end{equation*}
where $r \sim t(\nu,  \zeta^2{I}_n)$ was sampled independently from a student's \textit{t}-distribution. Here we considered $\rho \in \{0.1, 0.4, 0.7\}$ and $\nu \in \{3, 5, 10\}$ for error distribution and $\zeta^2 = \{0.1,0.2, \dots, 0.8\}$, $\eta^2_k = 0.3$, and an exponential decay for the signal component of the first $k$ eigenvalues. The true dimension was $k=10$.

Both non-normal error distribution and correlated features are expected to induce a change in the spectrum of the observed eigenvalues while the total amount of variance in $X$ (i.e.\ the sums of squared singular values) remains constant after standardization ($=n(m-1)$). This shift in the eigenvalue spectrum can occur in the first few eigenvalues in the case of correlated features or towards the middle of the eigenvalues in the case of the \textit{t}-distributed error, modifying the true SNR and thus making the estimation of effective rank more difficult.

\begin{figure}[H]
\centering
\includegraphics[scale=0.125]{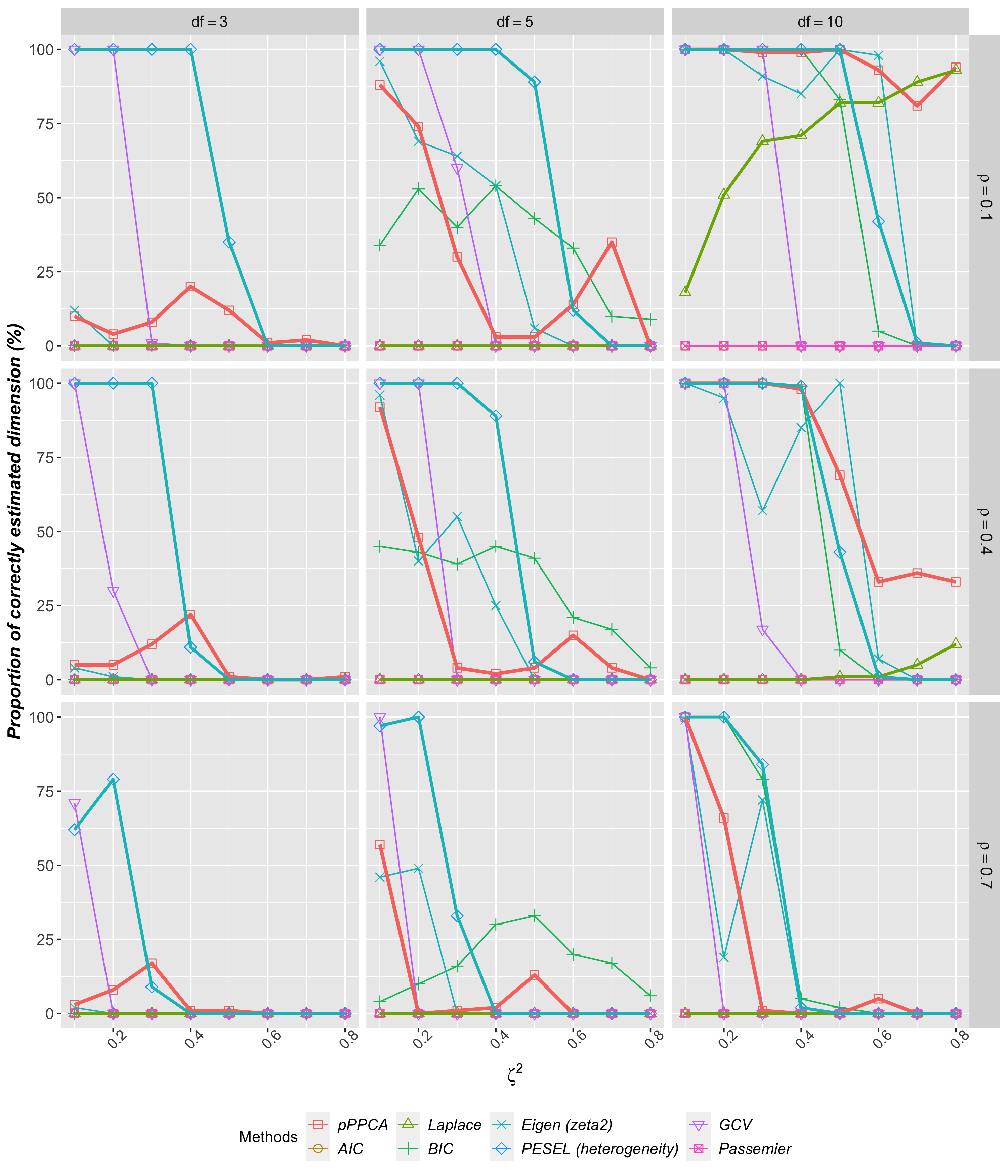}
\caption{Proportion of correctly estimated dimension over 100 replicates as a function of $\zeta^2$ assuming an exponential decay in the first $k$ squared singular values under non-normality.} \label{fig:S3}
\end{figure}

The fat tails and correlation in the error distribution present a challenge to \textit{Laplace} as the criterion were derived based under the normal distribution (Figure~\ref{fig:S3}). Naturally, a poor estimation of the residual variance $\zeta^2$ leads to an incorrect estimation of $k$, which affects all methods under comparison. Indeed, the residual variance would impact the estimated dimension through its relative size to the explained variance. A biased $\zeta^2$ estimate has a direct impact on the estimated dimension provided that the signal remains the same: a smaller $\hat{k}$ is expected for an upward biased $\zeta^2$ estimates, while a larger $\hat{k}$ is  expected for a downward biased $\zeta^2$ estimate. In the setting considered here, $n < m$, a data rich case as our interest is in the samples rather than features, the MLE estimator of $\zeta^2$ is consistent and approximately unbiased. However, when $n > m$, there will be a downward bias that requires the use of a biased corrected estimator~\cite{passemier2014estimation,passemier2017estimation}. 

Meanwhile, we observed that the estimated dimension decreased when the correlation structure was pronounced and wide spread (with the most impact on the first $k$ singular values). On the other hand, if the correlation were moderate, meaning the size and the number of variables involves are small, then the impact was mostly through the bottom $n-k$ singular values. But in this case, the impact was much smaller as the estimate $\zeta^2$ is a function of the average $n-k$ last singular values, while the top $k$ singular values are less influenced.

All methods except \textit{GCV}, \textit{PESEL (heterogeneity)}, \textit{Eigen (zeta2)} and \textit{pPPCA}, failed completely at identifying the true $k$, when $\rho = 0.7$ alone (Supplementary Figure 2) and $\nu = 3$ alone (Supplementary Figure 3). The proposed \textit{pPPCA} is more robust to correlation than distribution with fat tails as can be seen from the adequate performance under $\text{df}= 10$ and $\rho \le 0.4$ for a range of SNRs, suggesting tolerance for moderated correlated normal data (Figure~\ref{fig:S3}). On the other hand, \textit{PESEL (heterogeneity)} is more competitive when $df = 3$ and $\rho \ge 0.7$. Though \textit{GCV} is not as strong compared to \textit{PESEL (heterogeneity)}, the observed performance is expected considering it approximates a cross-validation criterion and does not depend on the underlying error distribution.

\section{Application to microarray gene expression data}\label{s:app_geneExp}

Large-scale gene expression data over multiple tissues have made it possible for scientists to study the global structure of expression profiles~\cite{lukk2010global} and extract biologically relevant information. It has been reported that linear projections of expression data have intrinsically low dimensions, but higher than previously thought~\citep{heimberg2016low,lenz2016principal,ding2018interpretable}. Here we apply the proposed method to a heterogeneous gene expression dataset to inform the effective rank.

\subsection{NCI60 Data}

This data contained gene expression measured across 9 types of human cancer cell lines~\citep{Ross:2000aa}, and has been recently profiled using microarray technology at $m=41,000$ gene probes~\citep{Liu:2010aa}. The pre-processed data were obtained from the European Bioinformatics Institute database and a total of $n = 60$ samples were analysed after removing 65 duplicated cell line samples (Table~\ref{t:nci60_class}). 

As only 30-40\% of genes are expected to expressed in each tissue~\citep{su2002large}, a standard variance filter was applied to remove gene probes with variance lower than their 10\% percentile value. In many cases, the excessively large variance corresponds to expression with bi-modal or even multi-modal distribution, and thus we removed gene probes with variance above 95\% percentile. The sizes of variance filters roughly correspond to 0.2 and 5.8 on the $log_{2}$ scale, which reduced the number of gene probes from $m=41,000$ to $m = 34,850$. See Supplementary Figure 4 for a summary of the sample and gene variance, as well as gene-based skewness and kurtosis prior to filtering. For each gene probe, the expression values were further standardized across samples to have a sample mean of zero and variance of 1. The sample eigenvalues were calculated based on the singular values of $X$ after standardization to be $\hat{\lambda}_i = \frac{\hat{d}^2_i}{m}$.  

\begin{table}[H]
\centering
\begin{tabular}{lc}
\hline
\hline
Tissue of origin & Number of samples\\
\hline
Breast & 6 \\
Central nervous system & 6 \\
Colon & 10 \\
Leukaemia & 7 \\
Melanoma & 11 \\
Non-Small Cell Lung  & 8 \\
Ovarian & 7 \\
Prostate & 2 \\
Renal & 9 \\
\hline
\end{tabular}
\caption{NCI 60 cell line classes.}\label{t:nci60_class}
\end{table}

\subsection{Data analysis}

\begin{figure}[H]
\includegraphics[scale=0.25]{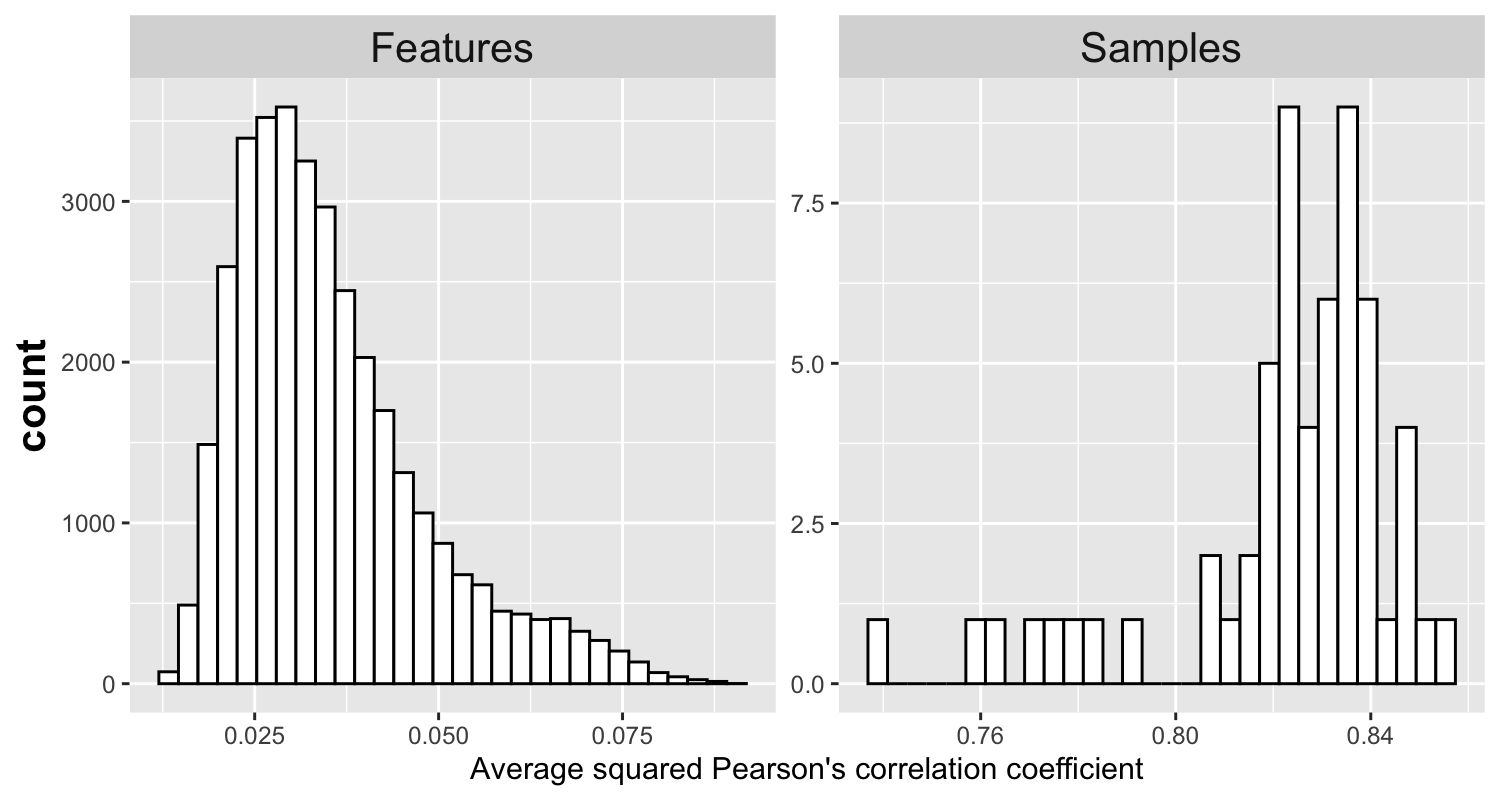}
\caption{Averaged squared Pearson's correlation coefficient for each sample or feature.}\label{fig:nci60_average_correl}
\end{figure}

Since correlation in both rows and columns is expected of gene expression data, we assessed the burden of such correlation using the averaged squared Pearson's correlation coefficient for each gene or sample (Figure~\ref{fig:nci60_average_correl}). In addition, gene expression distribution can be notoriously non-normal, with more than 50\% of gene features exhibiting heavy tails, skewness, and even multiple modes~\cite{de2020shape,marko2012non}. For a given dataset, we compared results on both the standardized data and those undergoing a rank-based inverse normal transformation for each gene feature. For alternative methods, only the most sensible estimate from a class of methods was reported, i.e.\ the value closest to the reported number of cell lines. Note that the reported results are exploratory in nature and had not been rigorously validated in terms of their biological interpretation nor clinical relevance.

As a follow-up analysis, we first estimated the dimension for the melanoma cell line alone since it had the highest number of samples (Table~\ref{t:nci60_class}), and then increased the number of samples by introducing additional cell line groups one by one, in the order of decreasing sample sizes per cell line (colon, renal, etc.). We hope the trend in estimated dimension as the data dimension increased can shed light on the structure of microarray data as they become increasingly more heterogeneous.

\subsection{Results}

\begin{figure}[H]
\includegraphics[scale=0.125]{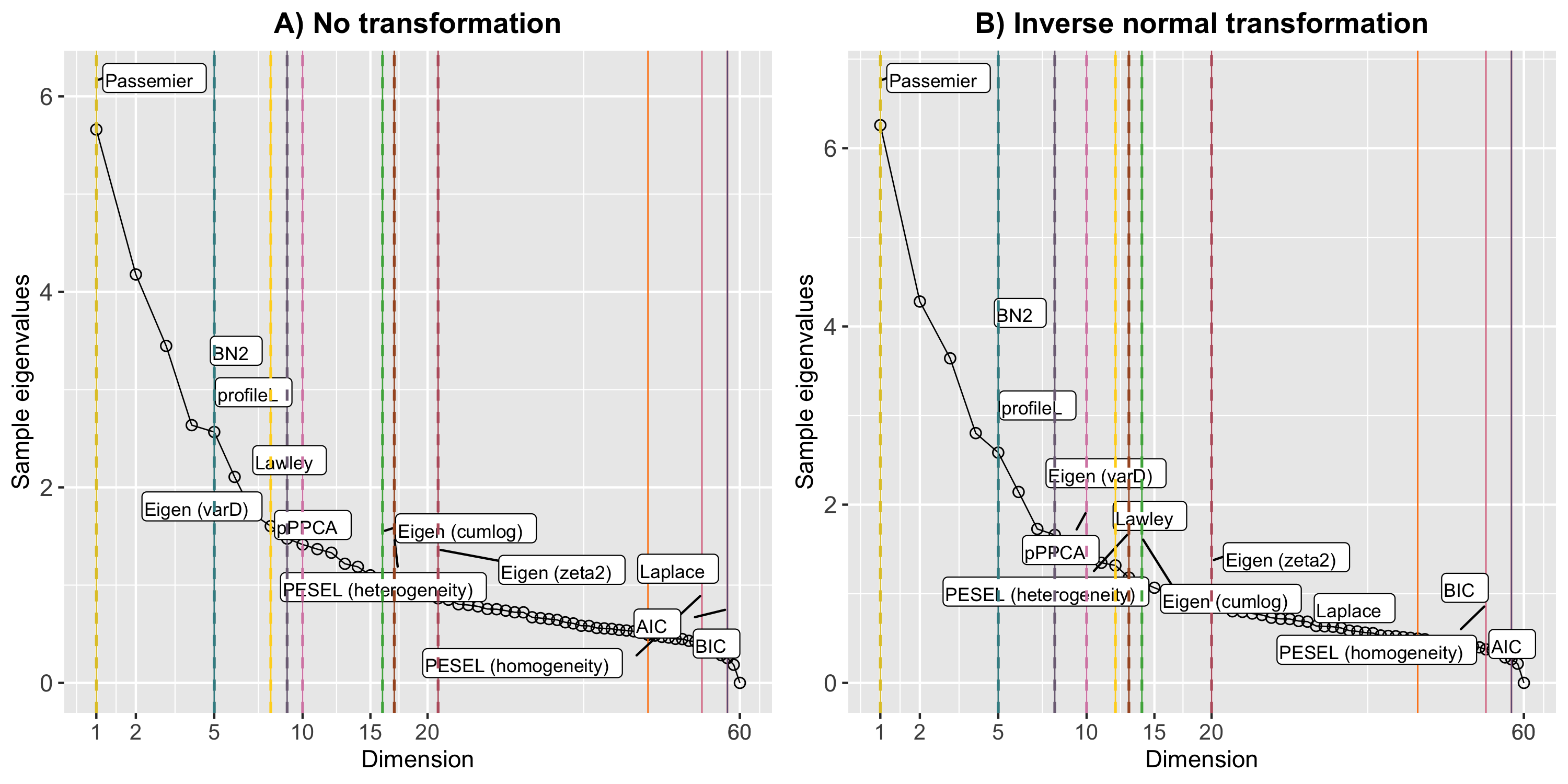}
\caption{Estimated effective rank by each method with respect to the sample eigenvalue scree plot.}\label{fig:nci60}
\end{figure}

There was no visible difference in the sample eigenvalues for data irrespective of a rank-based inverse normal transformation: in both cases we observed a smooth decay with no clear elbow (Figure~\ref{fig:nci60}). The penalized approach estimated $\tilde{k} = 10$ for both the standardized and the transformed data, suggesting robustness to non-normal features of the data. By design, empirical methods that are sensitive to the presence of a gap also gave similar estimates, for example, \textit{profileL}, \textit{GCV}, \textit{Lawley}, and elbow based approaches. Notably, \textit{GCV}, \textit{Lawley}, and the best of the elbow approach are in agreement with our penalized approach (Table~\ref{t:nci60}), giving estimates roughly in line with the number of cancer cell lines ($k=9$). On the other hand, model-based methods, such as \textit{AIC}, \textit{BIC}, and \textit{Laplace} were unable to gave sensible estimates. In particular, many overwhelmingly identified the boundary points at around $k=n-1$ or $k=1$. This observation agrees with Minka's comments in~\cite{minka2001automatic} that Bayesian methods do not perform well when data deviated from a reasonable level of normality and when the last $n-k$ sample eigenvalues decay faster than expected under the model, a result of either severe non-normality or the true eigenvalues of the last $n-k$ principal directions not being constant. The performance of \textit{PESEL (homogeneity)} seemed to suggest the later is more likely as it had shown fairly good performance under non-normality in simulations.

\begin{table}[H]
\centering
\begin{tabular}{lcc}
\toprule
Methods						& No transformation	 &  Inverse normal transformation \\
\midrule
Estimated dimension 					& 		 	& 		\\
{\textbf{pPPCA}} 	  				& 10			&	10	\\
  \cline{2-3}
AIC \								&	58 		& 58	\\
BN \									&   5		& 5	\\
BIC \							    &	58		& 58	\\
Best elbow approach\ 				&	9		&  8  \\
GCV\									&   10	 	& 12 	\\
Laplace\ 							&	46 		& 44 	\\
Lawley\ 								&	8 		& 12  \\
PESEL (heterogeneity) \				&	17		& 13 	\\
PESEL (homogeneity) \				&	54		& 54 	\\
Passemier\							&   58		& 58	 \\
ProfileL\							&   5		& 5	\\
\bottomrule
\end{tabular}
\caption{Estimated dimension of the NCI60 dataset.}\label{t:nci60}
\end{table}

\begin{figure}[H]
\includegraphics[scale=0.16]{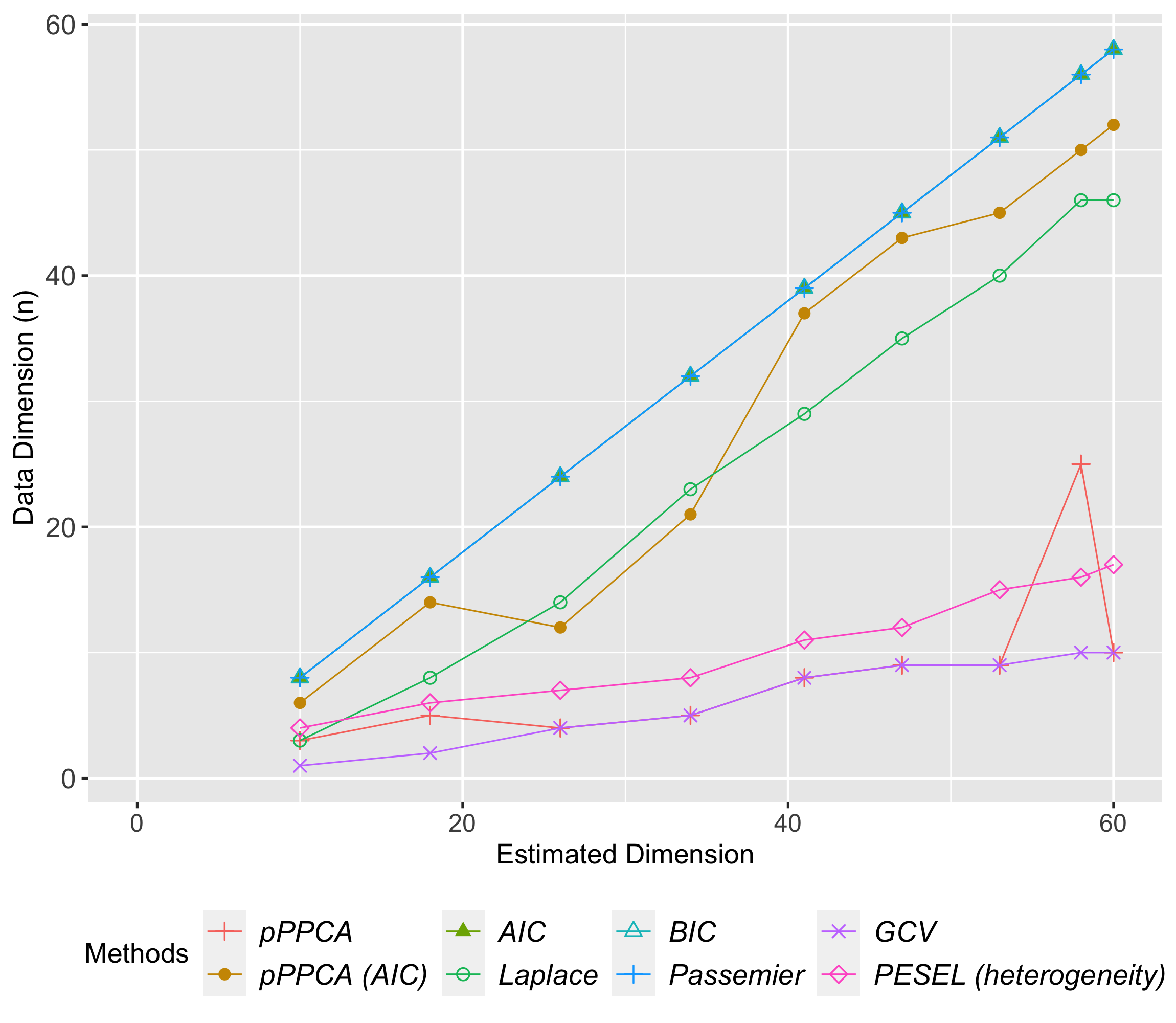}
\caption{Estimated effective rank with respect to increasing numbers of cell lines by each method.}\label{fig:NCI60_n}
\end{figure}

Since the expression data are heterogeneous coming from multiple cell lines, we sought to examine the data dimension as a function of increasing data complexity. Figure~\ref{fig:NCI60_n} reveals that the dimension increased with with additional cell line being included in the data.    

\section{Concluding remarks}

Both Bayesian methods and penalized approaches are often linked to improved prediction performance as a result of internally choosing the more parsimonious model. Here we compared their performance on the non-supervised learning of data dimension. Rather than an out-of-sample criterion, the estimation of dimension is very much ``in sample'' as we are primarily interested in the representation of this particular dataset and do not expect it to generalize beyond very homogeneous populations. 

The comprehensive simulation design covered a wide range of theoretical and realistic data scenarios, focusing on the impact of SNR, patterns of eigenvalue spectrum, relative sizes of $m$ and $n$, and correlated and non-normal error. The proposed \textit{pPPCA} strikes a balance between capturing the ``gap'' in the top sample eigenvalues via the voting strategy as well as modelling the error variance via a likelihood penalization. Thus, its complementary performance to the approximated Bayesian posterior likelihood and ``elbow'' based approaches is unsurprising. This also explains its good performance when data deviated from the independence assumption, an advantage in applications where one might be uncertain of the characteristics of the data generating process. 

Even though the proposed method was not the ``best'' in every scenario, its overall performance was competitive. Irrespective of other simulation conditions, it has good performance for large $k$ as the penalty on the estimated dimension is mostly driven by $\log{(\zeta^2)}$, which favours a model that is more flexible than preferred by Bayesian model selection. 

Supported by the application results, we recommend applying \textit{pPPCA} to explore the dimension of gene expression data when there is a good separation between signal and noise, and proper data transformation applied. As a possible follow-up analysis, the data could be better modelled assuming $\tilde{k}$ distinct error variance parameters using a generalized factor analysis model. Though in an exploratory analysis, the assumption of isotropic error covariance should suffice as a first step to identify the hidden dimension.

%

%

\bibliographystyle{tfnlm}
\bibliography{pPPCA}

\appendix

\section{Penalized PPCA and Minka's criterion using Laplace's method} \label{ss:pppca_minka}

Denote $Lambda = \text{diag}(\lambda_1, \dots, \lambda_{n})$, where $\lambda_i =  \eta^2_i + \zeta^2$. Intriguingly, if we use the following priors as suggested in Minka~\cite{minka2001automatic}:
\begin{align}
p(U, \Lambda, V, \zeta^2)& =p(\zeta^2)p(U)p(R)\Pi_{i=1}^k p(\lambda_i);\\
p(\zeta^2) &= \frac{1}{\Gamma(\alpha (n-k)/2)\zeta^2}\Big( \frac{\beta(n-k)}{2\zeta^2}\Big)^{\alpha (n-k)/2} \exp \Big(-\frac{\beta (n-k)}{2\zeta^2} \Big);\\
p(U) & = 2^{-k}\Pi_{i=1}^k \Gamma((n-i+1)/2)\pi^{-(n-i+1)/2};\\
p(\lambda_i) &= \frac{1}{\Gamma(\alpha/2)\lambda_i}\Big( \frac{\beta}{2\lambda_i}\Big)^{\alpha/2} \exp \Big(-\frac{\beta}{2\lambda_i} \Big);
\end{align}
and maximize the posterior with respect to $(\lambda_i, \zeta^2)$ at the maximum likelihood of $U$, we have:
\begin{align}
\hat{\zeta^2}(\alpha, \beta) &= \frac{m}{m-1+\alpha}\hat{\zeta}^2_k + \frac{\beta}{m(n-k)};\\
\hat{\lambda}_i(\alpha, \beta) &= \frac{m}{m-1+\alpha}\hat{\lambda}_i + \frac{\beta}{m}.
\end{align}

The final approximated Laplace evidence removed any terms that do not depend strongly on $k$ and simplified assuming $m$ is large and $(\alpha, \beta)$ are small.

The penalized MLE from our proposed penalty function~\eqref{eqn:pen} then corresponds to the hyperparameter values of: 
\begin{align}
\alpha 	&= 1-\frac{k}{n-k}\delta,\\
\beta	&= 0;
\end{align}
while the second penalty function~\eqref{eqn:pen2} corresponds to:
\begin{align}
\alpha 	&= 0,\\
\beta	&= \frac{k}{n-k}\delta,
\end{align}
that coincides with a prior using Levy distribution.

\section{Estimation of effective rank via penalized PPCA}\label{ss:pPPCA_details}

\subsection{Lemmas}

\begin{lemma}\label{lemma_21}
Consider a sample $X \in \mathbb{R}^{n \times m}$ with each column following a multivariate Gaussian distribution $\mathcal{N}(0, WW^{T} + \zeta^2{I})$. Suppose $W$ has rank $k$ and further, the sample covariance matrix of $X^{T}$ is positive semi-definite. Then, the penalized maximum log-likelihood at each fixed $q \in \{1, \dots, n-1\}$ is a smooth function of $\tilde{\delta}$ on the interval $(0,1/q-1/n)$ and is monotonically decreasing on
\begin{equation}
(0,(1/q-1/n)[1-\hat{\zeta}^{2}_q]), \label{eqn:delta_range}
\end{equation}
where $\hat{\zeta}^{2}_q = (\sum_{i=q+1}^{n} \hat{\lambda}_i)/(n-q)$.
\end{lemma}

Since the difference of two smooth functions is still a smooth function, the monotonicity of $l_p(q;\tilde{\delta}) -  l_p(q+1;\tilde{\delta})$ and $l_p(q;\tilde{\delta}) -  l_p(q-1;\tilde{\delta})$ can be established with respect to $\tilde{\delta}$.

\begin{lemma}\label{lemma_22}
Consider $\tilde{\delta} \in G(q+1)$, where
\begin{equation}
G(q+1) = \Big( 0, \frac{1}{n}\frac{(n-q-1)(\hat{\lambda}_{q+1}-\hat{\zeta}^{2}_{q+1})}{(q+1)\hat{\lambda}_{q+1} + (n-q-1)\hat{\zeta}^{2}_{q+1}}\Big).
\end{equation}
Then, for any fixed $q \in \{2, 3, \dots, n-2\}$, $l_p(q;\tilde{\delta}) -  l_p(q+1;\tilde{\delta})$ is a monotonically increasing and concave function of $\tilde{\delta} \in G(q+1)$ and $l_p(q;\tilde{\delta}) - l_p(q-1;\tilde{\delta})$ is a monotonically decreasing and convex function of $\tilde{\delta} \in G(q+1)$.
\end{lemma}

Since $l_p(q; \tilde{\delta}_o)$ is a discrete function of $q$, the maximum can be at either the boundary points or interior points. Considering exclusively the interior points, for some $q \in \{2, \dots, n-2\}$ to be the maximizer of $l_p(q; \tilde{\delta}_o)$ given $\tilde{\delta}_o$, $l_p(q; \tilde{\delta}_o) - l_p(q-1; \tilde{\delta}_o) > 0$ and $l_p(q; \tilde{\delta}_o) - l_p(q+1;  \tilde{\delta}_o) > 0$ constitute a necessary but not sufficient condition. With the additional condition that $l_p(q; \tilde{\delta}_o)$ monotonically increases $\forall q < k$ and monotonically decreases $\forall q > k$, the condition~\ref{condsAppend} becomes necessary and sufficient. The following Lemma proves the sufficiency of the condition that guarantees the true dimension $k$ to be the maximizer for some $\tilde{\delta}_o \in \cup_{q} G(q+1)$.

\begin{lemma}\label{lemma_23}
Assume the same notation from Lemma~\ref{lemma_22}. For $k \in \{2, \dots, n-2\}$, there exists $\tilde{\delta}_o \in \cup_{q} G(q+1)$ such that ${k} =\operatorname{argmax}_{q}  l_p(q; \tilde{\delta}_o)$ if and only if
\begin{equation}
\begin{cases}
    l_p(q; \tilde{\delta}_o) - l_p(q-1; \tilde{\delta}_o)  > 0 \\
    l_p(q; \tilde{\delta}_o) - l_p(q+1;  \tilde{\delta}_o)  > 0 \label{condsAppend}.
\end{cases}
\end{equation}
\end{lemma}

It is convenient to define the sets that satisfy~\ref{condsAppend} for $2 \le q \le n-2$:
\begin{equation}
\Delta_{q} = (a_{q}, b_{q}) \subset \cup_{q} G(q+1),
\end{equation}
where
\begin{equation}
a_{q}= \text{min} \left\{ \tilde \delta \in \cup_{q} G(q+1); \;  l_p(q;\tilde{\delta}) - l_p(q + 1;\tilde{\delta})  > 0 \right\}
\end{equation}
and
\begin{equation}
b_{q}= \text{max}\left\{\tilde \delta \in \cup_{q} G(q+1); \; l_p({q};\tilde{\delta}) - l_p(q - 1;\tilde{\delta})  > 0 \right\}.
\end{equation}

\begin{remark}
Note that an interesting result from Lemma~\ref{lemma_23} showing $\hat{\lambda}_{q+1} = 1$ to be a sufficient condition for $l_p(q;\tilde{\delta}) - l_p(q+1;\tilde{\delta}) > 0$ on $\tilde\delta \in G(q+1)$. This coincides with with Kaiser's rule for selecting $k$ as the number of PCs to retain. Notice that as $m \to \infty$,
$$\hat{\lambda}_{k+1}  \to \lambda_{k+1} = \zeta^2,$$
while the observed $\hat{\zeta}^2_k <  1$, then $\hat{\lambda}_{k} > \hat{\lambda}_{k+1} = 1$ provides strong evidence that the true $\eta^2_k = \lambda_k - \zeta^2 > 0$.
\end{remark}

\begin{lemma}\label{lemma_24}
Consider $$\Delta_{q}=\left \{\tilde{\delta} \in G(q+1); \; \mbox{conditions~\eqref{condsAppend} are satisfied} \right \} = (a_q, b_q),$$ whenever $a_q$ exists. Then $\Delta_{q}$ can be approximated by $(u_a(q), u_b(q)) \subset \Delta_q \subset G(q+1)$, where $u_a(q)$ represents an upper bound for $a_k$, and $u_b(q)$ a lower bound for $b_q$, such that $b_q/a_q > \frac{u_b(q)}{u_a(q)}$.
\end{lemma}

\begin{remark}
Clearly, $\frac{b_q}{a_q} > \frac{u_b(q)}{u_a(q)}$ holds. If the ratio $\frac{u_b(q)}{u_a(q)}$ converges as $m \to \infty$, the ratio asymptotically reflects the amount of evidence for each possible dimension $q$. Thus, if $\frac{u_b(q)}{u_a(q)}$ were the largest for $q=k$, then a majority-voting strategy for estimating $k$ is viable.
\end{remark}

\begin{lemma}\label{lemma_25}
Suppose $k$ is the true rank of $W$, then as $m \to \infty$,
\begin{itemize}
  \item $u_b(k)/u_a(k) \to \infty$ in probability
  \item $|u_b(q) - u_a(q)| \to 0$ in probability for $q > k$.
\end{itemize}
\end{lemma}

\begin{remark}
In theory, $u_b(k)/u_a(k)\to \infty$ in probability and the approximated ratio will be the largest as compared to other choices. However, in finite samples, the ratio $u_b(q)/u_a(q)$ for $q > k$ could also be quite large due to the numerical inaccuracy of the last $n-q$ sample eigenvalues as they approach the population values. In practice, the penalty tuning parameter $\tilde\delta$ needs to be calibrated such that $u_a(q)$ is not too close to 0.
\end{remark}

\begin{remark}\label{remark:221}
 The proof of Proposition~\ref{Proposition2} implies that given any $\tilde{\delta} = \tilde{\delta}_o$, a non-boundary maximizer of \eqref{eq:penloglik}, $k$, can be identified using the following conditions whenever $1 < q < k_\text{max}(\tilde\delta_o)$:
\begin{equation}
\begin{cases}
    l_p(q;\tilde{\delta}_o) - l_p(q + 1;\tilde{\delta}_o)  > 0; \\
    l_p(q;\tilde{\delta}_o) - l_p(q - 1;\tilde{\delta}_o)  > 0,
\label{cond}
\end{cases}
\end{equation}
where $k_\text{max}(\tilde\delta_o)$ is the maximum value for the search space that ensures $\tilde{\zeta}^2_q$ is well-defined given $\tilde{\delta} = \tilde{\delta}_o$. In other words, $l_p(q;\tilde{\delta}_o)$ first increases with $q \le k$ and then decreases with $q \ge k$, thus ensuring $k$ maximizes $l_p(q;\tilde{\delta}_o)$ over $q \in \{2, \dots, k_\text{max}(\tilde\delta_o)\}$.
\end{remark}

\begin{remark}\label{remark:222}
It is clear that $\Delta_{q}$ is an open interval for each $q$ as the penalized likelihood in~\eqref{eq:penloglik} is a continuous function of $\tilde\delta \in (0, 1/(q+1)-1/n)$ for any fixed $q$ (Lemma~\ref{lemma_21}). Following Lemmas~\ref{lemma_22} and~\ref{lemma_23}, for $q \ne q'$, $\Delta_q$ and $\Delta_{q'}$ are strictly non-overlapping sets. Therefore, the realized range for $\tilde{\delta}$ is the union of all sets $\cup_{q=2}^{n-2}\Delta_q \subset [0, 1-1/n)$. But because of the restriction embedded in \eqref{eq:sigmak} and \eqref{condsAppend}, we must have $\tilde{\delta} \in (0, 1/(q+1)-1/n)$ for each examined value of $q$. Consequently, the restriction imposes a relationship whereby $q$ is non-increasing in $a_q$ (or $u_a(q)$) and $b_q$ (or $u_b(q)$). For example, when $q = n-1$, it must be that $a_{n-1} = 0 < b_{n-1} < \frac{(n-1)^{-1}-n^{-1}}{1-\hat{\zeta}^2_{n-1}}$, while for $q = 1$, $ b_2 < a_1 < b_1 < (1-n^{-1})\{1-\hat{\zeta}^2_1\}$.
\end{remark}

\begin{remark}\label{remark:224}
For $q=1$ or $q=n-1$, $\Delta_{q}$ can be defined such that only one of~\eqref{condsAppend} is satisfied. It is clear that $\Delta_{q}$ is an open interval for each $q$ as the penalized likelihood function in \eqref{eq:penloglik} is a continuous function of $\delta \in G(q)$ for any fixed $q$. However, in this case, as $a_{n-1} = 0$ and $b_1$ is unbounded, results from Lemmas~\ref{lemma_24} and~\ref{lemma_25} no longer apply. Instead, a practical solution is to construct suitable probabilistic models for $q=0$ and $q=n$ such that the boundary points become interior points.
\end{remark}

\begin{remark}\label{remark:223} Since $a_q$ and $b_q$ are not analytically available, whenever possible, I obtained conservative upper and lower bounds for $\Delta_q$ using $u_a(q)$ and $u_b(q)$ such that $(u_a(q), u_b(q)) \subset \Delta_{q} $ (Lemma~\ref{lemma_24}). The proof of Lemma~\ref{lemma_25} also demonstrates that ${u_b(k)}/{u_a(k)} > 1$ so that $ (u_a(k), u_b(k)) \ne \varnothing$. Essentially, the number of votes provides a form of evidence for division between the first $q$ and last $n-q$ sample eigenvalues relative to the first $q-1$ and last $n-q+1$ or the first $q+1$ and last $n-q-1$.
\end{remark}

Lemmas~\ref{lemma_21},~\ref{lemma_22},~\ref{lemma_23},~\ref{lemma_24}, and \ref{lemma_25} together imply: 1) there exists $\tilde{\delta}_o \in {\Delta}_{k}$ such that \eqref{eq:penloglik} is maximized at $k$; 2) ${\Delta}_{k} = (a_{k}, b_{k})$ can be approximated by $(u_a(k), u_b(k)) \subset \Delta_{k}$, satisfying
\begin{eqnarray}
\lim_{m \to \infty} \frac{u_b(k)}{u_a(k)}&=& \infty, \\
\lim \limits_{m \to \infty} |u_b(q) - u_a(q)| &\to& 0, \mbox{ for } q >k ,\\
\lim_{m \to \infty} \frac{u_b(q)}{u_a(q)}&<& \infty, \mbox{ for }  q<k.
\label{intervals}
\end{eqnarray}

\subsection{Establishing the range of plausible tuning parameter values}\label{ss:range}

We first consider the search space for $k$. The smallest and the largest non-trivial choice for $k$ is 1 and $n-1$, respectively. Clearly, the largest possibly value that $k$ can take depends on the actual rank of the sample covariance. We define $n_{\text{max}} = \text{min} \{i: \hat{\lambda}_i < \kappa\}-1$ where $\kappa$ is a tolerance value that can be set arbitrarily low to prevent digits over-floating in standard software (e.g.\ $\kappa = 0.001$). This, in effect, removes numerical uncertainty in the inverse of sample eigenvalues. 

The construction of search grid is characterized by its range and the distance between adjacent grid values. Results from Lemma~\ref{lemma_25} suggest a possible construction, $\tilde{\delta}_1,\ldots, \tilde{\delta}_T$, using a sequence of $T$ equidistant points on $\log$ scale. To determine $\tilde{\delta}_1$ and $\tilde{\delta}_T$, we need to bound the minimum and the maximum of $\tilde{\delta}$ values such that 1 and $n_\text{max}$ are the maximizer of the penalized profile log-likelihood~\eqref{eq:penloglik}. Since the exact relationship between $q$ and $a_q, \; b_q$ is not analytically available, we rely on conservative bounds obtained via Taylor series approximations to specify $\tilde{\delta}_1$ and $\tilde{\delta}_T$. 

The maximum value $\tilde{\delta}_T$ is defined as the average of the two largest penalties that $\tilde{\zeta}^2_1 > \tilde{\zeta}^2_2$ and $\tilde{\zeta}^2_2 > \tilde{\zeta}^2_3$ hold, as above these values, the model is overwhelmed by the penalty and will always choose $k=1$. The minimum value $\tilde{\delta}_1$ is chosen to be the value given by $u_a(n_\text{max})$.

In practice, the boundary points $q=1$ and $q=n-1$ might be relevant, and have the interpretations of $n-1$ independent error or signal components, respectively. To curb the definition of a maximizer according to~\ref{condsAppend}, we propose to construct artificial boundary points for which the penalized profile log-likelihood is defined for $q = n$ or $q = n_\text{max}$ and $q = 0$.

Define $\hat{\lambda}_0 = \sum_{i=1}^{n} \hat{\lambda}_i $ and the error variance is then $\hat{\zeta}^2_0 = 1$. This model corresponds to $W = \mathbf{0}$ and thus $q=0$. By design, $l_p(q=0;\tilde\delta) = l_p(q=0)$ as there is no dimension to penalize. On the other hand, since $\hat{\zeta}_{n-1} = \hat{\lambda}_n$ and $l_p(q=n-1) = l_p(q=n)$, the construction must impose a small probabilistic component to the $q=n$ model (e.g.\ PCA) by introducing $\hat{\zeta}^2_n = \frac{1}{2}\hat{\lambda}_{n} = \frac{1}{2}\hat{\zeta}^2_{n-1}$. This model corresponds to a dimension that is between $q=n-1$ and $q=n$ and forces $l_p(q=n;\tilde\delta) = l_p(q=n) < l_p(q=n-1)$.

These artificially constructed boundary points makes it possible to select a maximum for the penalized maximum likelihood by choosing $q = 1$ or $q = n-1$ such that $l_p(q; \delta)- l_p(q-1; \delta) >0$ and $l_p(q; \delta) - l_p(q+1; \delta) > 0$.

\section{Alternative methods} \label{append:methods}

Details of the alternative methods considered in the simulation studies are included here and organized in alphabetical order.

\subsection{Akaike information criterion (\textit{AIC})}
The number of free parameters in the model is $nq + 1 - q(q-1)/2$ and the model with the smallest \textit{AIC} is selected:
\[
AIC(q) = -2l_p(q)+ nq + 1 - q(q-1),
\]
where $l_p(q)$ is defined as in~\eqref{eq:maxlk}.

\[
\hat{k} = \operatorname{arg\,min}_q AIC(q).
\]

\subsection{Bayesian information criterion (\textit{BIC})}
A simplification to the Laplace's method assuming $m\to \infty$ \citep{kass1995bayes}:
\begin{equation}
\log p(D|q) = -\frac{m}{2}\Big(\sum_{i=1}^q \log{\hat{\lambda}_i}\Big)  -\frac{m(n-q)}{2}\log \Big(\frac{\sum_{i=q+1}^n \hat{\lambda}_i}{n-q}\Big) -\frac{nq-(q+1)q/2+q}{2} \log{(m)}, \label{eqn:BICevidence}
\end{equation}
where any terms that do not depend on $m$ are dropped. It can be shown that this simplifies to the likelihood under a model subtracted by a multiple of the number of free parameters, which is the usual \textit{BIC} criterion $BIC(q) = -2l_p(q)+ \frac{nq + 1 - q(q-1)}{2}\log m$.

\[
\hat{k} = \operatorname{arg\,min}_q BIC(q).
\]

\subsection{Bai and Ng's criteria (\textit{BN})}

Bai and Ng~\citep{Bai2002} developed six different criteria via a selection of penalty functions involving both $m$ and $n$ to identify the number of factors, where the errors are allowed to be correlated. The inference was performed jointly on $(k, \zeta^2)$.

The three criteria applicable to PPCA models are:
\begin{equation}
\hat{k} = \operatorname{arg\,min}_q  V(q, \hat{F}^{q}) + q \hat{\zeta}^2_{BN} g_j(m, n),
\end{equation}
where $\hat{\zeta}^2_{BN} = \frac{1}{nm}\sum_{j=1}^{m}\sum_{i=1}^n\text{E}(x_{ij})^2$, $V(q, \hat{F}^{q}) = \frac{1}{nm} \sum_{j=1}^m\text{E}(X_j^{T}X_j)$ and the three penalty functions:
\begin{equation}
g_{1}(m, n) = \frac{m+n}{nm} \log{\Big(\frac{nm}{n+m} \Big)},
\end{equation}
\begin{equation}
g_{2}(m, n) = \frac{m+n}{nm} \log{\text{min}(n, m)},
\end{equation}
and
\begin{equation}
g_{3}(m, n) = \frac{\log{\text{min}(n, m)}}{\text{min}(n, m)}.
\end{equation}

Following the PPCA model, the criterion reduces to
\begin{equation}
\hat{k} =  \operatorname{arg\,min}_q \hat{\zeta}^2_q + q \hat{\zeta}^2_{k_o} g_j(m, n),
\end{equation}
where $k_o$ is the maximum number of PCs searched. Alternatively, the estimators $\hat{\zeta}^2_q$ can be replaced by the bias corrected estimators introduced in~\cite{passemier2017estimation}. Thus, giving a total of 6 criteria used for comparison. For $k_o$, I chose $\lfloor\frac{n}{2}\rfloor$ as it gave the best performance across scenarios.

\subsection{Empirical elbow approaches (\textit{Elbow})}

I have also included in the comparison a few empirical approaches designed to detect an ``elbow'' or a point of inflection in the scree plot produced by the sample eigenvalues:
\begin{enumerate}
\item
 The difference between log cumulative mean of the sample eigenvalues and the mean of the cumulative log sample eigenvalues (\textit{cumlog}), defined by
\[
\widehat{k}_{\text{cumlog}} = \operatorname{arg\,min}_q \log{\frac{\sum_{i=1}^q {\hat{\lambda}_i}}{q}} - \frac{1}{q}\sum_{i=1}^q \log{\hat{\lambda}_i},
\]

\item the variance of sample eigenvalues (\textit{VarD}), defined by
\begin{equation}
\widehat{k}_{\text{VarD}} = \operatorname{arg\,min}_q \frac{\sum_{i=1}^q \hat{\lambda}_i^2}{q} - \left (\frac{\sum_{i=1}^q \hat{\lambda}_i}{q} \right)^2,
\end{equation}
\item the adjacent sample eigenvalues (\textit{adjD}), defined by
\begin{equation}
\widehat{k}_{\text{adjD}} = \operatorname{arg\,min}_q \frac{\hat{\lambda}_q}{\hat{\lambda}_{q+1}},
\end{equation}
\item and a criterion based on the $\log$ of estimated error variance (\textit{log-var}), defined by 
\begin{equation}
\widehat{k}_{\text{log-var}} = \operatorname{arg\,min}_q (n-q)\log\hat{\zeta}_q^2.
\end{equation}
\end{enumerate}

\subsection{A general cross-validation criterion (\textit{GCV})}

This criterion is similar to the general cross-validation in regression to approximate the leave-one-out cross-validation, which is based on the relationship between prediction error and residual sum of squared via a weight matrix resulted from a projecting matrix. This enables a smoothing approximation to cross-validation criterion results in a general cross-validation (\textit{GCV}) criterion that is computationally advantageous:
\begin{equation}
\hat{k}_\text{GCV} = \operatorname{arg\,min}_q  \frac{m^2n \sum_{i = q+1}^n \hat{\lambda}_i}{[(m-1)n - mq - nq + q^2 + q]^2}.
\end{equation}
To produce optional results, data would be transposed if the number of observations were smaller than sample size.

\subsection{An approximation to the posterior likelihood using Laplace's method (\textit{Laplace})}
Laplace approximation \citep{minka2001automatic,hoyle2008automatic} assumes the dimension of the parameter space is constant. Thus, ${Z}$ is integrated out~\citep{minka2001automatic} and the resulting posterior likelihood is approximated using Laplace's method~\citep{kass1995bayes}, which requires the $\operatorname{arg\,max}$ of the parameters and the Hessian matrix at these values.

The log of the evidence is:
\begin{align}
\nonumber \log{p(D|q)} &  = \log{p(U)} -m/2\Big(\sum_{i=1}^q \log{\hat{\lambda}_i}\Big) - m(n-q)/2 \log{\Big(\frac{\sum_{i=q+1}^n \hat{{\lambda}}_i}{n-q}\Big)}   \\
\nonumber & + \frac{2nq-q^2+q}{4}\log(2\pi) - q/2\log{(m)} -1/2 \sum_{i=1}^q\sum_{j=i+1}^n\Big[\log(\frac{(\hat{\lambda}_i-\hat{\lambda}_j)^2}{\hat{\lambda}_i\hat{\lambda}_j}) + \log(m)\Big] \\
\nonumber & = -m/2\Big(\sum_{i=1}^q \log{\hat{\lambda}_i}\Big) - \frac{m(n-q)}{2} \log{\Big(\frac{\sum_{i=q+1}^n \hat{{\lambda}}_i}{n-q}\Big)} -1/2 \sum_{i=1}^q\sum_{j=i+1}^n\Big[\log\frac{(\hat{\lambda}_i-\hat{\lambda}_j)^2}{\hat{\lambda}_i\hat{\lambda}_j}\Big]\\
& +\frac{2nq-3q-q^2}{4}\log{(2)} - \frac{q^2 - 2nq +3q }{4}\log{(m)} + \sum_{i=1}^q\log{\Gamma(\frac{n-i+1}{2})}, \label{eqn:Laplceevidence}
\end{align}
where
\begin{equation}
\log{p(U)} = -q\log{(2)} + \sum_{i=1}^q\log{\Gamma(\frac{n-i+1}{2})} - \frac{2nq+q-q^2}{4}\log{(\pi)}
\end{equation}

\subsection{A hypothesis testing criterion for the equality of the last $n-k$ eigenvalues (\textit{Lawley})}

The null hypothesis is $H_o: \lambda_j = \lambda_{j+1} = \dots, = \lambda_n$ against the alternative hypothesis that at least one is not equal to the remaining eigenvalues. The test  statistic is given by \cite{lawley1956tests}:

\begin{equation}
\chi^2 = (n-j)c\log{\Big(\frac{\sum_{i=j}^n \hat{\lambda}_i}{n-j}\Big)} - \sum_{i=j}^{n} \log\hat{\lambda}_i
\end{equation}
where $$c = (n-j)-\frac{2(n-j) +1+2/(n-j)}{6} + \left(\frac{\sum_{i=j}^n \hat{\lambda}_i}{n-j}\right)^2\sum_{i=1}^j \left(\sum_{i=1}^j \hat{\lambda}_i- \frac{\sum_{i=j}^n \hat{\lambda}_i}{n-j}\right)^{-2},$$ 
with $\frac{(n-j)(n-j+1)}{2}-1$ degrees of freedom.

\subsection{A PEnalized Semi-integrated Likelihood (\textit{PESEL})}
The criteria proposed here~\citep{sobczyk2017bayesian} are inspired by BIC, which assumes the number of free parameters is independent of the number of observations, and clearly this is not always satisfied. The rationale is to integrate out some parameters from~\eqref{eq:generative_model}, either elements in ${Z}$ so the model does not depend on $m$ (i.e. $m \to \infty$) or to integrate out ${W}$ so the model selection does not depend on $n$ (i.e. $n \to \infty$). A total of four criteria are given under different asymptotics with respect to $n$ and $m$ while considering the first $k$ eigenvalues are equal (homogeneous) or different (heterogeneous).

\begin{itemize}
\item Fixed $m$ with $n \to \infty$: $\text{PESEL}_{n}$
	\begin{itemize}
		\item $\text{PESEL}_{n, \text{heter}}$ is equivalent to the BIC approximation in \cite{minka2001automatic}.

    	\begin{align}
		\nonumber\text{PESEL}_{n, \text{heter}}(q) = & \frac{-mn}{2}\log{(2\pi)} - \frac{n}{2}\sum_{j=1}^q log{(\hat{\lambda}_j}) - \frac{n(m-q)}{2}\log{(\hat{\zeta}^2_q)} \\
			&-\frac{mn}{2} -\log{(n)}\frac{mq - \frac{q(q+1)}{2} + q + \mathbf{m} + 1}{2}
	\end{align}

		\item $\text{PESEL}_{n, \text{homo}}$ assumes all PCs have the same variance (i.e. there is no dominant direction)
			\begin{align}
			\nonumber\text{PESEL}_{n, homo}(q) =& \frac{-mn}{2}\log(2\pi) - \frac{n\mathbf{q}}{2}log({\frac{\sum_{j=1}^q \hat{\lambda}_j}{q}}) - \frac{n(m-q)}{2}\log(\hat{\zeta}^2_q) \\
			&-\frac{mn}{2} -\log(n)\frac{mq - \frac{q(q+1)}{2} + q + \mathbf{1} + 1}{2}
			\end{align}
	\end{itemize}

\item Fixed $n$ with $m \to \infty$: $\text{PESEL}_{m}$

	\begin{itemize}
		\item $\text{PESEL}_{m, \text{heter}}$

			\begin{align}
			\nonumber\text{PESEL}_{m, heter}(q) =& \frac{-mn}{2}\log(2\pi) - \frac{m}{2}\sum_{j=1}^q log({\hat{\lambda}_j}) - \frac{m(n-q)}{2}\log(\hat{\zeta}^2) \\
			&-\frac{mn}{2} -\log(m)\frac{nq - \frac{q(q+1)}{2} + q + \mathbf{n} + 1}{2}
			\end{align}

		\item $\text{PESEL}_{m, \text{homo}}$
			\begin{align}
			\nonumber\text{PESEL}_{m, homo}(q) =& \frac{-mn}{2}\log(2\pi) - \frac{m\mathbf{q}}{2}log({\frac{\sum_{j=1}^q \hat{\lambda}_j}{q}}) - \frac{m(n-q)}{2}\log(\hat{\zeta}^2_q) \\
			&-\frac{mn}{2} -\log(m)\frac{nq - \frac{q(q+1)}{2} + n + \mathbf{1} + 1}{2}
			\end{align}
	\end{itemize}
\end{itemize}

\subsection{A bias-corrected criterion (\textit{Passemier})}

With the main asymptotic assumptions as follows:
 \begin{align}
 	n & \to \infty \\
 	m & \to \infty \\
 	c_n & = \frac{m}{n-1} \to c > 0,
 \end{align}
\cite{passemier2017estimation} proposed a plug-in estimator for $\zeta^2$ using a bias correction that depends on $q$:
\begin{equation}
\hat{\zeta}^2_{*} = \hat{\zeta}^2 + \frac{b(\hat{\zeta}^2)}{n-q}\hat{\zeta}^2\sqrt{2c_n}.
\end{equation}
where $b(\zeta^2) = \sqrt{c/2}\{q + \zeta^2\sum_{i=1}^k(1/\lambda_i)\}$.

Without the correction, the noise variance is expected to have a downward bias as $n$ increases relative to $m$. A consistent estimator for the true number of PCs ($k$) under $m >> n$ is given, where it is assumed that $k << n$. The proposed criterion to select $k$ requires a tuning parameter to be chosen and the default value is $b = 0.05$ for each $q$:
\begin{equation}
\hat{k}_\text{Passemier} = \operatorname{arg\,min}_q \hat{\zeta}^2_{q*} + q\hat{\zeta}^2_{k_o}\frac{(c_n+2\sqrt{c_n})(1+m/n^{1+b})}{n},
\end{equation}
where $k_o$ is the maximum number of PCs searched. In preliminary simulation results, I observed that $b$ needs to be bigger than the default $0.05$ to obtain the correct estimate in some cases, especially for the more difficult cases with smaller SNR. Thus, besides the default value of $0.05$, I also included the 95\% and 5\% quantile values of $\{\hat{\lambda}_i\}_{i=1,\dots, n}$, and the best results from these choices are reported.

\subsection{A profile likelihood-based criterion (\textit{ProfileL})}

\cite{zhu2006automatic} proposed a simple profile likelihood-based criterion to detect the ``elbow'' by separating the first $q$ and last $n-q$ sample eigenvalues under the following models:
\begin{equation}
\hat{\lambda}_j \sim \mathcal{N}(\mu_1, \gamma); \quad j \le q
\end{equation}
and
\begin{equation}
\hat{\lambda}_j \sim \mathcal{N}(\mu_2, \gamma);  \quad j > q.
\end{equation}

The profile likelihood evaluates the evidence for a change-point by maximizing:
\begin{equation}
pL(q) = \sum_{j=1}^q\log\mathcal{N}(\hat{\lambda}_j | \mu_1(k), \gamma(q)) + \sum_{j=k+1}^n\log\mathcal{N}(\hat{\lambda}_j | \mu_2(k), \gamma(q)),
\end{equation}
where $\mu_1(q)$ and $\mu_2(q)$ are estimated by the mean sample eigenvalues in each partition separated by $q$, while $\gamma(q)$ is given by a pooled estimate using all sample eigenvalues.

\end{document}